\documentclass{aa}
\usepackage{graphicx}
\usepackage{bm}
\usepackage{times}
\usepackage{amsmath}
\usepackage{amssymb}
\usepackage{natbib}
\usepackage{color}
\usepackage{mathrsfs}
\usepackage[colorlinks,allcolors=blue]{hyperref}
\usepackage{url}
\usepackage{multirow}
\bibpunct{(}{)}{;}{a}{}{,}
\graphicspath{{./fig/}{./png/}}
\newcommand{\fff}{{\bm f}}

\newcommand{\uuu}{{\bm u}}

\newcommand{\xxx}{{\bm x}}

\newcommand{\AAA}{{\bm A}}

\newcommand{\BBB}{{\bm B}}
\newcommand{\mBBB}{\overline{\bm B}}

\newcommand{\FFF}{{\bm F}}

\newcommand{\JJJ}{{\bm J}}

\newcommand{\UUU}{{\bm U}}
\newcommand{\mUUU}{\overline{\bm U}}

\newcommand{\gggg}{{\bm g}}

\newcommand{\mUUvp}{\overline{U}_\varpi}

\newcommand{\mUUp}{\overline{U}_\phi}

\newcommand{\mUUz}{\overline{U}_z}

\newcommand{\mBBz}{\overline{B}_z}

\newcommand{\mBBvp}{\overline{B}_\varpi}

\newcommand{\mBBp}{\overline{B}_\phi}

\newcommand{\bcdot}{\bm\cdot}
\newcommand{\Eq}[1]{Eq.~(\ref{#1})}
\newcommand{\Equ}[1]{Equation~(\ref{#1})}

\newcommand{\Eqs}[2]{Equations~(\ref{#1}) and (\ref{#2})} 
\newcommand{\Equs}[2]{Equations~(\ref{#1}) to (\ref{#2})}

\newcommand{\EQ}{\begin{equation}}
\newcommand{\EN}{\end{equation}}
\newcommand{\EQA}{\begin{eqnarray}}
\newcommand{\ENA}{\end{eqnarray}}
\newcommand{\brac}[1]{\langle #1 \rangle}
\newcommand{\pd}{\partial}

\newcommand{\mean}[1]{\overline{#1}}

\newcommand{\cP}{c_{\rm P}}
\newcommand{\cV}{c_{\rm V}}
\newcommand{\cs}{c_{\rm s}}

\newcommand{\urms}{u_{\rm rms}}

\newcommand{\brms}{B_{\rm rms}}
\newcommand{\orms}{\omega_{\rm rms}}

\newcommand{\Prot}{P_{\rm rot}}

\newcommand{\Ma}{{\rm Ma}}

\newcommand{\chiSGSo}{\chi_{\rm SGS}^{(1)}}

\newcommand{\Co}{{\rm Co}}

\newcommand{\Pe}{{\rm Pe}}

\newcommand{\PraSGS}{{\rm Pr}_{\rm SGS}}

\newcommand{\Pm}{{\rm Pm}}

\newcommand{\Ra}{{\rm Ra}}

\newcommand{\Rey}{{\rm Re}}

\newcommand{\ReM}{{\rm Re}_{\rm M}}
\newcommand{\Ro}{{\rm Ro}}

\newcommand{\Ta}{{\rm Ta}}

\newcommand{\taucool}{\tau_{\rm cool}}

\newcommand{\mOm}{\mean{\Omega}}

\newcommand{\mvOm}{\mean{\bm\Omega}}

{}

\newcommand{\Rgas}{{\cal R}}

\newcommand{\Omsi}{\Omega^{\rm sim}}
\newcommand{\Omssi}{\Omega_\odot^{\rm sim}}

\newcommand{\Rsi}{R^{\rm sim}}

\newcommand{\gsi}{g^{\rm sim}}

\newcommand{\Lratio}{L_{\rm ratio}}
\newcommand{\FFFrads}{{\bm F}_{\rm rad}}
\newcommand{\FFFSGSs}{{\bm F}_{\rm SGS}}

\newcommand{\mLenth}{\mean{L}_{\rm enth}}
\newcommand{\mLconv}{\mean{L}_{\rm conv}}
\newcommand{\mLrad}{\mean{L}_{\rm rad}}
\newcommand{\mLkin}{\mean{L}_{\rm kin}}
\newcommand{\mLvisc}{\mean{L}_{\rm visc}}

\newcommand{\mLcool}{\mean{L}_{\rm cool}}
\newcommand{\mLheat}{\mean{L}_{\rm heat}}
\newcommand{\Ekin}{E_{\rm kin}}
\newcommand{\Emag}{E_{\rm mag}}

\def\onethird{{\textstyle{1\over3}}}

\def\onehalf{{\textstyle{1\over2}}}

\newcommand{\Figa}[1]{Fig.~\ref{#1}}
\newcommand{\Figas}[2]{Figs.~\ref{#1} and \ref{#2}}
\newcommand{\Figu}[1]{Figure~\ref{#1}}

\newcommand{\Table}[1]{Table~\ref{#1}}

\newcommand{\Appendix}[1]{Appendix~\ref{#1}}

\begin{document}

\authorrunning{K\"apyl\"a}
\titlerunning{Star-in-a-box simulations of fully convective stars}

   \title{Star-in-a-box simulations of fully convective stars}

   \author{P. J. K\"apyl\"a
          \inst{1,2}
          }

   \institute{Georg-August-Universit\"at G\"ottingen, Institut f\"ur 
              Astrophysik, Friedrich-Hund-Platz 1, D-37077 G\"ottingen,
              Germany
              email: \href{mailto:pkaepyl@uni-goettingen.de}{pkaepyl@uni-goettingen.de}
   \and
              Nordita, KTH Royal Institute of Technology and Stockholm
              University, SE-10691 Stockholm, Sweden}

\date{\today}

\abstract{Main-sequence late-type stars with masses less than $0.35
  M_\odot$ are fully convective.
   }%
   {The goal is to study convection, differential rotation, and
      dynamos as functions of rotation in fully convective stars.
   }%
   {Three-dimensional hydrodynamic and magnetohydrodynamic numerical
     simulations with a star-in-a-box model, where a spherical star is
     immersed inside of a Cartesian cube, are used. The model
     corresponds to a $0.2M_\odot$ main-sequence M5 dwarf. A range of
     rotation periods ($\Prot$) between 4.3 and 430 days is explored.
   }%
   {The slowly rotating model with $\Prot = 430$ days produces
     anti-solar differential rotation with a slow equator and fast
     poles, along with predominantly axisymmetric quasi-steady
     large-scale magnetic fields. For intermediate rotation ($\Prot =
     144$ and $43$ days) the differential rotation is solar-like (fast
     equator, slow poles), and the large-scale magnetic fields are
     mostly axisymmetric and either quasi-stationary or cyclic. The
     latter occurs in a similar parameter regime as in other numerical
     studies in spherical shells, and the cycle period is similar to
     observed cycles in fully convective stars with rotation periods
     of roughly 100 days. In the rapid rotation regime the
     differential rotation is weak and the large-scale magnetic fields
     are increasingly non-axisymmetric with a dominating $m=1$
     mode. This large-scale non-axisymmetric field also exhibits
     azimuthal dynamo waves.
   }%
   {The results of the star-in-a-box models agree with simulations of
     partially convective late-type stars in spherical shells in that
     the transitions in differential rotation and dynamo regimes occur
     at similar rotational regimes in terms of the Coriolis (inverse
     Rossby) number. This similarity between partially and fully
     convective stars suggests that the processes generating
     differential rotation and large-scale magnetism are insensitive
     to the geometry of the star.
   }%

  \maketitle


\section{Introduction}

The existence of fully convective low-mass stars has been predicted by
theoretical stellar models at least since the 1950s
\citep[e.g.][]{1958ApJ...127..363L}. Current stellar structure and
evolution models indicate that stars with masses less than about
$0.35M_\odot$ are fully convective
\citep[e.g.][]{1997A&A...327.1039C,2000ARA&A..38..337C}. However,
until recently no direct observational evidence could be attached to
this transition. A recent analysis of GAIA observations indicates a
gap in the main sequence at spectral type M3 which is perhaps the most
direct observational evidence of transition to full convection to date
\citep{2018ApJ...861L..11J}. The gap in the Hertzsprung--Russell
diagram near the transition to full convection is presumed to be
associated with an instability related to $^3$He burning in the core
of the star, such that the core undergoes periods of radiative and
convective phases before settling to a final either convectively
stable or unstable state depending on the mass of the star
\citep[e.g.][]{vSP12,2018A&A...619A.177B,2021ApJ...907...53F}. Furthermore,
in the final stable configuration, stars just on the partially
convective side of the transition have a substantial (roughly 40 per
cent of stellar radius) radiative core, implying a discontinuity in
stellar structure around $M=0.35M_\odot$.

The transition to full convection is important from the point of view
of stellar dynamo theory because of an ongoing debate regarding the
origin of the solar large-scale magnetic field \citep[e.g.][and
  references therein]{2020LRSP...17....4C}. This debate revolves
around the importance of the tachocline, a layer of strong shear just
below the convection zone (CZ), and that of the convection zone
itself. In simplified terms, in the distributed dynamo scenario the
magnetic fields are generated throughout the CZ and the tachocline is
unimportant \citep[e.g.][]{KR80,Br05,KKT06,PK11}, while in the
flux-transport models \citep[e.g.][]{DC99,2014ApJ...782...93H} the
roles of the two are opposite. In reality, however, stellar dynamos
are likely to have ingredients from both scenarios.

The fact that there are no tachoclines in fully convective stars makes
these objects highly interesting in view of dynamo theory. If M star
magnetism would show a discontinuity at the transition to full
convection, one could argue that also the type of dynamo changes
across the transition. However, no conclusive observational evidence
for such discontinuity has been detected so far
\citep[][]{2008MNRAS.390..545D,2008MNRAS.390..567M,2010MNRAS.407.2269M,2021A&ARv..29....1K},
such that most activity indicators appear to be continuous ac-cross the
transition to full convection \citep[see,
  however,][]{2015ApJ...812....3W,2020ApJ...891..128M}. For example,
the X-ray luminosity of fully convective M stars falls in line with
partially convective stars
\citep{2016Natur.535..526W,2018MNRAS.479.2351W}.

Observations of M stars shows a rich variety of possible magnetic
states \citep[see][and references
  therein]{2021A&ARv..29....1K}. Zeeman-Doppler Imaging (ZDI) results
of early-type (partially convective) M dwarfs tend to have relatively
weak multipolar fields while stars around the transition to full
convection often have strong dipole-dominated configurations
\citep[][]{2008MNRAS.390..545D}. Intriguingly, rapidly rotating
late-type M dwarfs can have either configuration, suggesting a
possibility of bistable dynamo states
\citep[e.g.][]{2013A&A...549L...5G}. The ZDI results of fully
convective stars have not yet revealed any cyclicity, for example in
terms of polarity reversals of the large-scale field. The lack of
sufficiently long data series is a likely contributor to
this. However, longer term observations of other activity indicators,
such as chromospheric emission, of a nearby slowly rotating
($\Prot\approx90$~days) fully convective star Proxima Centauri
indicate an activity cycle of around seven years
\citep[][]{2016A&A...595A..12S,2017MNRAS.464.3281W,Damassoeaax7467,2021MNRAS.500.1844K}.
Another recent study found evidence of an activity cycle of roughly
5~years for the fully convective star Ross 128 with rotation period of
more than 100 days \citep[][and references
  therein]{2019A&A...628L...1I}. Interestingly, an activity cycle of
similar length has also been reported recently from a rapidly rotating
fully convective star Gl~729 with $\Prot\approx 3$~days
\citep[e.g.][]{2020arXiv201110391I}.

Another aspect that makes low mass stars particularly interesting is
that detecting Earth-sized planets on orbits suitable for life around
M stars is much easier than for, for example, solar mass stars. This
is because M stars are less massive than the Sun, and due to their low
luminosity the habitable planets must have close orbits, leading to a
much larger radial velocity signal
\citep[e.g.][]{2014PNAS..11112641K}. However, as noted above, M stars
are very often magnetically active which leads to effects such as
starspots and Zeeman broadening which can hinder radial velocity
measurements. The magnetic activity of the host star is also of
significance for the exoplanets such that they can be subject to
violent flares and coronal mass ejections. Therefore, understanding
convection and magnetic activity of M stars is of great importance in
the search for possible life-supporting exoplanets.

Numerical simulations of fully convective stars are challenging due to
the fact that the convection zone (CZ) spans the entire radius of the
star, implying strong density contrast and greatly varying spatial and
temporal scales. Furthermore, the Mach numbers in the cores of fully
convective stars are very low leading to a short acoustic timestep. A
further complication arises due to the singularity of commonly used
spherical polar coordinates at the centre of the star. Most
three-dimensional simulations of fully convective stars to date
\citep[e.g][]{2008ApJ...676.1262B,2011IAUS..271...69B,2013A&A...549L...5G,YCMGRPW15,2016ApJ...833L..28Y,2020ApJ...902L...3B}
have been made using anelastic equations which bypass the acoustic
time step constraint. Furthermore, most of these simulations operate
in spherical shells and the innermost core of the star is not modeled
due to the coordinate singularity \citep[see,
  however,][]{2020ApJ...902L...3B}. Although it can be argued that the
omission of a small core spanning at most a few per cent of stellar
radius is likely to be insignificant for the global dynamics,
additional boundary conditions need to be imposed around the central
void. An exception is the study of \cite{DSB06} who used a
star-in-a-box approach \citep[see
  also][]{2002AN....323..213F,2004A&A...423.1101D,2019JPhCS1225a2020W,2020arXiv200108452M}
where a spherical star is embedded into a Cartesian cube. This
approach allows the full star, including the centre, to be
modeled. However, this comes at a cost of including the corners of the
box that are unimportant for the interior dynamics of the star. This
also implies that the spatial resolution of the star-in-a-box
simulations is necessarily lower than in corresponding spherical
shells models.

In the current study the star-in-a-box model of \cite{DSB06} is
refined with developments from convection simulations in other
contexts. This includes using heat conductivity based on Kramers
opacity law \citep[e.g.][]{2000gac..conf...85B,2017ApJ...845L..23K}
and a subgrid-scale entropy diffusion that enables more realistic
luminosities. Furthermore, one of the principal aims of the current
study is to compare the results from the star-in-a-box models of fully
convective stars with corresponding results obtained in spherical
shells. Another point of interest is to compare the transitions of
differential rotation and dynamo solutions as functions of rotation in
fully convective stars. Numerical studies of partially convective
stars indicate a transition from anti-solar (slow equator, fast poles)
to solar-like (fast equator, slow poles) differential rotation
\citep[e.g.][]{GYMRW14,KKB14} around the rotation rate where the
advection and Coriolis forces are of the same order of magnitude, that
is Coriolis (inverse Rossby) number unity. Furthermore, corresponding
dynamo solutions progress from predominantly axisymmetric and
quasi-steady solutions for slow rotation
\citep[e.g.][]{2017A&A...599A...4K,2018ApJ...863...35S}, through
oscillatory axisymmetric fields at intermediate rotation
\citep[e.g.][]{Gi83,KMB12,NBBMT13}, to non-axisymmetric azimuthal
dynamo waves for rapid rotation
\citep[e.g.][]{CKMB14,YGCR15,2018A&A...616A.160V}. A set of
simulations with rotation periods in the range from 4.3 to 430 days
are presented to study if, and where, such transitions in differential
rotation and dynamos occur also in models of fully convective stars.

\section{The model} \label{sect:model}

The model is based on the star-in-a-box model presented in
\cite{DSB06}. A star of radius $R$ is embedded into a Cartesian cube
with a side half-length of $H=1.1R$ where all coordinates $(x,y,z)$
run from $-H$ to $H$. Motions and temperature fluctuations are damped
for $r > R$ where $r =\sqrt{x^2+y^2+z^2}$ is the spherical radius. The
following set of equations of magnetohydrodynamics (MHD) are solved:
\begin{eqnarray}
\frac{\pd \AAA}{\pd t} &=& \UUU \times \BBB - \eta\mu_0\JJJ, \label{equ:indu} \\
\frac{D \ln \rho}{Dt} &=& - \bm\nabla \bm\cdot \UUU, \label{equ:conti} \\
\frac{D\uuu}{D t} &=& -\bm\nabla \Phi -\frac{1}{\rho}(\bm\nabla p - \bm\nabla \bm\cdot 2 \nu \rho \bm{\mathsf{S}} + \JJJ \times \BBB)\nonumber \\ && \hspace{4.2cm} - 2\ \bm\Omega \times \UUU + \fff_d, \label{equ:mom} \\
T \frac{D s}{D t} &=& -\frac{1}{\rho} \left[\bm\nabla \bm\cdot \left(\FFF_{\rm rad} + \FFF_{\rm SGS}\right) + {\cal H} - {\cal C} \right] \nonumber \\ && \hspace{4.2cm} + 2 \nu \bm{\mathsf{S}}^2 + \mu_0 \eta \JJJ^2, \label{equ:entro}
\end{eqnarray}
where $\AAA$ is the magnetic vector potential, $\UUU$ is the velocity,
$\BBB=\bm\nabla\times\AAA$ is the magnetic field,
$\JJJ=\bm\nabla\times\BBB/\mu_0$ is the current density, $\mu_0$ is
the permeability of vacuum, $\eta$ is the magnetic diffusivity, $D/Dt
= \pd/\pd t + \UUU \bm\cdot\bm\nabla$ is the advective derivative,
$\rho$ is the fluid density, $\Phi$ is the gravitational potential,
$p$ is the pressure, $\nu$ is the kinematic viscosity,
$\bm{\mathsf{S}}$ is the traceless rate-of-strain tensor with
\begin{eqnarray}
\mathsf{S}_{ij} = \onehalf (U_{i,j} + U_{j,i}) - \onethird \delta_{ij} \bm\nabla\bm\cdot\UUU,
\end{eqnarray}
where the commas denote differentiation and $\delta_{ij}$ is the
Kronecker delta. $\bm\Omega = (0,0,\Omega_0)$ is the rotation rate of
the star, $\fff_d$ is a damping function, $T$ is the temperature, $s$
is the specific entropy, $\FFFrads$ is the radiative flux, $\FFFSGSs$
is the subgrid-scale entropy flux, and ${\cal H}$ and ${\cal C}$
describe heating and cooling, respectively.

The gas obeys an ideal gas equation of state with $p = \Rgas \rho T$,
where $\Rgas=\cP - \cV$ is the universal gas constant, and $\cP$ and
$\cV$ are the heat capacities at constant pressure and volume,
respectively. Effects of degeneracy in the equation of state are
neglected. The gravitational potential $\Phi$ is fixed in space and
time, and set up such that it corresponds to an isentropic hydrostatic
state of an M5 star \citep[see Appendix~A of][]{DSB06} with
\begin{eqnarray}
\Phi(r) = -\frac{GM}{R} \frac{a_0 + a_2 r'^2 + a_3 r'^3}{1 + b_2 r'^2 + b_3 r'^3 + a_3 r'^4},
\end{eqnarray}
where $G$ is the gravitational constant, $M$ is the mass of the star,
$r' = r/R$ and $a_0 = 2.34$, $a_2 = 0.44$, $a_3 = 2.60$, $b_2 = 1.60$,
and $b_3 = 0.21$.

The term $\fff_d$ describes damping of flows exterior to the star
according to
\begin{eqnarray}
\fff_d = -\frac{\UUU}{\tau_{\rm damp}} f_{\rm e}(r),
\end{eqnarray}
where $\tau_{\rm damp}$ is a damping timescale, and where $f_{\rm
  e}(r)$ is given by
\begin{eqnarray}
f_{\rm e}(r) = \frac{1}{2} \left(1 + \tanh \frac{r-r_{\rm damp}}{w_{\rm damp}} \right),
\label{equ:udamp}
\end{eqnarray}
where $r_{\rm damp} = 1.03R$ and $w_{\rm damp} = 0.03R$. The damping
timescale, $\tau_{\rm damp} \approx 1$~day, is short in comparison to
all other relevant timescales in the system.

Radiation is taken into account though the diffusion
approximation. The radiative flux is given by
\begin{eqnarray}
\FFFrads = - K \bm\nabla T,
\end{eqnarray}
where
\begin{eqnarray}
K(\rho,T) = K_0 (\rho/\rho_0)^{a-1} (T/T_0)^{b+3}. 
\end{eqnarray}
Here $a=-1$ and $b=7/2$ are chosen corresponding to the Kramers
opacity law that describes the average opacity for bound-free and
free-free transitions \citep[e.g.][]{carroll2013introduction}. The
Kramers opacity law was first used in convection simulations by
\cite{2000gac..conf...85B}, but has been adopted more widely only
recently
\citep{2017ApJ...845L..23K,2019A&A...631A.122K,2020GApFD.114....8K,2021A&A...645A.141V}.

The use of Kramers heat conductivity entails that the diffusion of
entropy fluctuations depends strongly on density and temperature such
that the thermal diffusivity, $\chi(\xxx,t)=K(\xxx,t)/\cP
\rho(\xxx,t)$, varies several orders of magnitude as a function of
height \citep[e.g.][]{2019A&A...631A.122K}. Thus it is numerically
desirable to introduce subgrid-scale (SGS) entropy diffusion that does
not contribute to the net energy transport, but which damps
fluctuations near grid scale. This is implemented via an SGS entropy
flux \citep[e.g.][]{2015JPlPh..81e3904R}
\begin{eqnarray}
\FFFSGSs = - \chiSGSo \rho \bm\nabla s',
\end{eqnarray}
where
\begin{eqnarray}
s' = s - \brac{s}_t,
\label{equ:sfluct}
\end{eqnarray}
are the fluctuations of the entropy, and where $\brac{s}_t(\xxx,t)$ is
a running temporal mean of the specific entropy with an averaging time
interval of roughly $75$~days. This differs from commonly adopted SGS
models in that the fluctuation that the SGS term acts on,
\Equ{equ:sfluct}, is not based on a difference to a spatial average
\citep[e.g.][]{2020GApFD.114....8K}, or to a fixed reference state
\citep[e.g.][]{2008ApJ...676.1262B,GDW12}. This choice has a physical
and computational justification: spatially averaged means practically
always assume certain symmetries of the solution (for example,
spherical or axisymmetry) but even the actual mean state might not
have these symmetries if, for example, a strong low-$m$
non-axisymmetric magnetic field is present. Furthermore, time
averaging is computationally advantageous because no global
communication is not required. However, a caveat is that if the
entropy field acquires numerical artifacts they will also propagate to
$\mean{s}_t$. This was encountered in some of the most rapidly
rotating cases where hyperdiffusion proportional to $\nabla^6
\brac{s}_t$ was applied to ensure the smoothness of the time average.

The heating term ${\cal H}$ is a parameterization of nuclear energy
production in the core of the star and it is given by a normalized
Gaussian profile
\begin{eqnarray}
{\cal H}(r) = \frac{L_{\rm sim}}{(2\pi w_L^2)^{3/2}}
\exp\left(-\frac{r^2}{2w^2_L} \right),
\end{eqnarray}
where $L_{\rm sim}$ is the imposed luminosity in the simulation and
$w_L = 0.162R$ is the width of the Gaussian, see Fig.~1 of
\cite{DSB06}.

Finally, the term ${\cal C}$ in the entropy equation models radiative
losses above the stellar surface with a cooling term
\begin{eqnarray}
{\cal C}(\xxx) = \rho \cP \frac{T(\xxx) - T_{\rm surf}}{\taucool} f_{\rm e}(r),
\end{eqnarray}
where $\taucool = \tau_{\rm damp}$ is a cooling timescale, and
$f_{\rm e}$ is the same function as in \Eq{equ:udamp} with $r_{\rm
  cool} = r_{\rm damp}$ and $w_{\rm cool} = w_{\rm damp}$.

\subsection{Simulation strategy, units, and control parameters}

The use of fully compressible formulation of the MHD equations means
that using realistic stellar parameters leads to prohibitively long
Kelvin--Helmholtz timescale
\begin{eqnarray}
  \tau_{\rm KH} = \frac{GM^2}{RL},
\end{eqnarray}
where $L$ is the luminosity of the star, and a short time step (in
comparison to the advective time) due to the low Mach number. This is
addressed by using a much higher luminosity in the models than in the
real star. This is described by the luminosity ratio $\Lratio=L_{\rm
  sim}/L$ between the simulation and the star. The enhanced luminosity
leads to convective velocity enhancement according to $u_{\rm conv}
\propto \Lratio^{1/3}$. However, it is advantageous to reinterpret the
convective velocities produced in the simulation as the physical
velocity. This entails re-scaling the sound speed $\cs$ (temperature)
proportional to $\Lratio^{-1/3}$ ($\Lratio^{-2/3}$). Furthermore, the
Kelvin--Helmholtz time scales as $\tau_{\rm KH} \propto
\Lratio^{-1}$. This implies that the acoustic timestep $\delta t_{\rm
  ac}$ increases proportional $\Lratio^{1/3}$ and the gap between
$\delta t_{\rm ac}$ and $\tau_{\rm KH}$ reduces by a cumulative factor
of $\Lratio^{4/3}$. This procedure enables $\tau_{\rm KH}$ to be
resolved given that $\Lratio$ is sufficiently large. The main effect
of this is that the Mach number and relative fluctuations of
thermodynamic quantities are enhanced which, however, scale with known
power laws as functions of $\Lratio$ \citep{2020GApFD.114....8K}. The
main physical difference is that overshooting at the interfaces of
radiative and convective layers is also enhanced. Recent numerical
experiments suggest that the enhancement of overshooting is much
weaker than that of the convective velocities
\citep{2019A&A...631A.122K}. This aspect is, however, unimportant in
the fully convective setups considered here.

The use of an enhanced luminosity requires also that the radiative
conductivity $K$ and the surface cooling ${\cal C}$ are increased by
the same factor $\Lratio$ to preserve the same temperature gradient in
a hydrostatic (non-convecting) model. This is done by enhancing $K_0$
and decreasing $\tau_{\rm cool}$ correspondingly. In principle, the
kinematic viscosity and magnetic diffusivity should also be scaled
with a factor $\Lratio^{1/3}$ to preserve realistic values of the
Reynolds and P\'eclet numbers. In practice, however, the diffusivities
in the simulations are always much higher than the real physical
values. Thus the smallest values of $\nu$ and $\eta$ admitted by the
grid resolution are used. The scaling of diffusivities is important if
simulations with different $\Lratio$ are compared which is not the
case in the present study \citep[see,
  however,][]{2019A&A...631A.122K,2020GApFD.114....8K}.

The units of length, time, density, entropy, and magnetic field are
\begin{eqnarray}
&& \hspace{1cm} [x] = R,\ \ \ [t] = \sqrt{R^3/GM}, \ \ \ [\rho] = \rho_0, \label{equ:units1} \\  && \hspace{1.5cm} [s] = \cP,\ \ \ [B] = \sqrt{\mu_0 \rho_0}\ [x]/[t], \label{equ:units2}
\end{eqnarray}
where $\rho_0$ is the initial value of the density at the centre of
the star. The results are often quoted in physical units such that the
quantities on the rhs of \Eqs{equ:units1}{equ:units2} use values from
literature (see below).

The following input parameters define the models. The dimensionless
luminosity ${\cal L}$ is given by \citep{DSB06}
\begin{eqnarray}
{\cal L} = \frac{L_{\rm sim}}{\sqrt{G^3M^5/R^5}},
\end{eqnarray}
where ${\cal L} = 5.5 \cdot10^{-5}$ for the runs discussed in
the present study. This is three orders of magnitude smaller than in
\cite{DSB06} but still much higher than in an main-sequence M5
dwarf. The amount of density stratification is determined by the
dimensionless pressure scale height at the surface
\begin{eqnarray}
\xi_0 = \frac{{\cal R} T_{\rm surf}}{GM/R},
\end{eqnarray}
where $T_{\rm surf} = T(R)$. The current simulations have $\xi_0
\approx 0.062$ which results in an initial density contrast of roughly
50 which is an order of magnitude greater than in the simulations of
\cite{DSB06}.

Stellar parameters for an M5 dwarf used by \cite{DSB06} are
$M=0.21M_\odot$, $R=0.27R_\odot$, $L=0.008L_\odot$, $T_{\rm eff}
=4000$~K, and $\rho_0 = 1.5\cdot 10^5$~kg~m$^{-3}$. Therefore, ${\cal
  L_{\rm M5}} = 2.4 \cdot 10^{-14}$ and $\xi_{\rm M5} = 2.2 \cdot
10^{-4}$. Thus the simulations have a much higher luminosity and lower
density stratification than in reality. With ${\cal L} = 5.5
\cdot10^{-5}$ the luminosity ratio between the simulation and an M5
star is $\Lratio = {\cal L}/{\cal L_{\rm M5}} \approx 2.1 \cdot
10^{9}$. Thus the velocities are greater by a factor $\Lratio^{1/3}
\approx 1280$ such that the rotation rate of the star needs to be
enhanced by the same factor to model a realistic rotational influence
on the flow. This means that the centrifugal force would be
unrealistically large in comparison to the acceleration due to gravity
if it were explicitly included \citep[see][]{2020GApFD.114....8K}.
Nevertheless, the Mach number $\Ma=\urms/\cs$ of the convective flows
in the current simulations is on average $0.1$ such that the time step
is dominated by the sound speed at the centre of the
star. Furthermore, the Kelvin--Helmholtz time is $\tau_{\rm KH}
\approx 1.84 \cdot 10^4 \sqrt{R^3/GM}$, which corresponds to 375 years
in physical units. While this is still achievable with current
computational resources, lower values of $\Lratio$ quickly lead to
infeasible computational demands due to the combined effect of lower
$\Ma$ and longer $\tau_{\rm KH}$. Thus the chosen value of $\Lratio$
should be considered as a compromise between realism and computational
feasibility.

\begin{table*}[t]
\centering
\caption[]{Summary of the runs.}
  \label{tab:runs1}
       \vspace{-0.5cm}
      $$
          \begin{array}{p{0.05\linewidth}ccccccccccccc}
          \hline
          \hline
          \noalign{\smallskip}
          Run & P_{\rm rot} [\mbox{days}] & \Ta [10^8] & \urms [{\rm m}/{\rm s}] & \brms [\mbox{kG}] & \Co & \Co^{(\omega)} & \Ra_{\rm t} & \Rey & \Pe & \ReM & \tau_{\rm sim} [\mbox{years}] & \mbox{DR} & \mbox{grid} \\
          \hline
          HD1   &   -   &   -   &   18  &   -   &    -   & -    &  6.5 \cdot 10^5 & 189  &  37  &   -  &  302 & -  & 288^3 \\
          \hline
          RHD1  &  433  & 0.16  &   17  &   -   &   0.6  &  0.1 &  2.3 \cdot 10^6 & 181  &  36  &   -  &   58 & \mbox{anti-solar} & 288^3 \\
          RHD2  &  144  &  1.4  &   15  &   -   &   1.9  &  0.2 &  1.7 \cdot 10^6 & 159  &  31  &   -  &   37 & \mbox{solar-like} & 288^3 \\
          RHD3  &   43  &   16  &   15  &   -   &   6.4  &  0.9 &  4.8 \cdot 10^6 & 159  &  31  &   -  &   36 & \mbox{solar-like} & 288^3 \\
          RHD4  &   14  &  144  &  9.0  &   -   &    31  &  3.0 &  1.3 \cdot 10^7 &  96  &  19  &   -  &  220 & \mbox{solar-like} & 288^3 \\
          RHD5  &  4.3  & 1600  &  5.7  &   -   &   166  &   11 &  7.9 \cdot 10^7 &  61  &  12  &   -  &  209 & \mbox{solar-like} & 288^3 \\
          \hline
          MHD0  &   -  &   -  &   18  &  \mbox{no dynamo}  &  -  & - &  6.5 \cdot 10^5  & 189  &  37  &  95  &  87 & - & 288^3 \\
          MHD1  &  433  & 0.16  &   14  &  11  &   0.7  &  0.1 &  1.2 \cdot 10^6 & 145  &  29  &  72  &  143 & \mbox{anti-solar} & 288^3 \\
          MHD1h &  433  &  1.4  &   14  &  12  &   0.7  &  0.1 &  3.6 \cdot 10^6 & 293  &  58  & 146  &   79 & \mbox{anti-solar} & 576^3 \\
          MHD2  &  144  &  1.4  &   14  & 4.5  &   2.0  &  0.3 &  1.7 \cdot 10^6 & 152  &  30  &  76  &  107 & \mbox{solar-like} & 288^3 \\
          MHD3  &   43  &   16  &   10  &  10  &   9.1  &  1.2 &  3.2 \cdot 10^6 & 111  &  22  &  55  &  134 & \mbox{solar-like} & 288^3 \\
          MHD3h &   43  &   64  &   10  &  12  &   9.4  &  1.4 &  8.0 \cdot 10^6 & 215  &  43  & 107  &   83 & \mbox{solar-like} & 576^3 \\
          MHD4  &   14  &  144  &  5.8  &  13  &    49  &  3.6 &  7.8 \cdot 10^6 &  62  &  12  &  31  &  201 & \mbox{solar-like} & 288^3 \\
          MHD5  &  4.3  & 2500  &  2.6  &  19  &   366  &   24 &  1.1 \cdot 10^7 &  34  &   6  &  17  &  302 & \mbox{solar-like} & 288^3 \\
          MHD5h &  4.3  & 10^4  &  2.5  &  23  &   375  &   14 &  3.7 \cdot 10^7 &  67  &  13  &  33  &  149 & \mbox{solar-like} & 576^3 \\
          \hline
          \end{array}
          $$ \tablefoot{All simulations have
            $\mathcal{L}=5.5\cdot10^5$. The diffusion coefficients
            in the lower resolutions ($288^3$) runs are $\nu =
            2.8 \cdot 10^6$~m$^2$/s, $\chiSGSo = 1.4 \cdot
            10^7$~m$^2$/s, and $\eta = 5.6 \cdot 10^6$~m$^2$/s, in all
            cases except in run MHD5 where $\nu = 2.2 \cdot
            10^6$~m$^2$/s, $\chiSGSo = 1.1 \cdot 10^7$~m$^2$/s, and
            $\eta = 4.4 \cdot 10^6$~m$^2$/s. In runs MHD1h and
              MHD3h $\nu = 1.4 \cdot 10^6$~m$^2$/s, $\chiSGSo = 7.0
              \cdot 10^6$~m$^2$/s, and $\eta = 2.8 \cdot
              10^6$~m$^2$/s, whereas in run MHD5h $\nu = 1.1 \cdot
              10^6$~m$^2$/s, $\chiSGSo = 5.5 \cdot 10^6$~m$^2$/s, and
              $\eta = 2.2 \cdot 10^6$~m$^2$/s. Run MHD0 is the same as
              run HD1 but with magnetic fields included. The initial
            value of $\chi = 2.0 \cdot 10^3$~m$^2$/s at the centre of
            the star. $\tau_{\rm sim}$ is the length of the
            statistically steady part of the simulations and the last
            column indicates the sense of the differential rotation
            (DR).}
\end{table*}

The surface gravity for the M5 star is
\begin{eqnarray}
g = \frac{GM}{R^2} = \frac{0.21}{(0.27)^2} \frac{GM_\odot}{R^2_\odot} \approx 2.9 g_\odot \approx 750\ \frac{\rm m}{{\rm s}^2},
\end{eqnarray}
where $g_\odot = 260$~m~s$^{-2}$. Thus the conversion factor between
the rotation rate in the simulations and physical units is
\citep[cf.\ Appendix~A of][]{2020GApFD.114....8K}
\begin{eqnarray}
\Omsi = \Lratio^{1/3} \left(\frac{\gsi}{g} \frac{R}{\Rsi}
\right)^{1/2} \Omega,\label{equ:Omegaconv}
\end{eqnarray}
where $g$, $R$, $\Omega$ are the surface gravity, stellar radius, and
rotation rate in physical units and quantities with superscript `sim'
in simulation (code) units. For example, assuming an M5 star rotating
at the solar rotation rate $\Omega_\odot = 2.7 \cdot
10^{-6}$~s$^{-1}$ yields
\begin{eqnarray}
\Omssi = 1.73 \left(\frac{\gsi}{\Rsi} \right)^{1/2}.
\end{eqnarray}
Rotation rates corresponding to rotation periods $\Prot =
2\pi/\Omega_0$ between 4.3 and 430 days are considered in the present
study.

The Taylor number is given by
\begin{eqnarray}
\Ta = \frac{4 \Omega_0^2 R^4}{\nu^2}.
\end{eqnarray}
The remaining control parameters are the SGS and magnetic Prandtl
numbers
\begin{eqnarray}
\PraSGS = \frac{\nu}{\chiSGSo},\ \ \ \Pm = \frac{\nu}{\eta}.
\end{eqnarray}
In the current study $\PraSGS = 0.2$ and $\Pm = 0.5$ are used
throughout. The radiative diffusion $\chi=K/\rho \cP \ll \chiSGSo$
within the star in the cases considered here is such that the
corresponding Prandtl number is $\gg 1$, see \Table{tab:runs1}.  The
value of $\chi$ is artificially capped to $\chi<0.1\chiSGSo$ because
it would otherwise become very large in the corners of the box due to
the very low density there and prohibitively limit the timestep of the
simulations.

Impenetrable and stress-free boundary conditions are imposed for the
flow and the magnetic field is assumed to be perpendicular to the
boundary. Temperature is assume symmetric accross the exterior
boundaries of the box and vanishing second derivative is assumed for
the density.

The simulations were run with the {\sc Pencil
  Code}\footnote{\href{https://github.com/pencil-code/}{https://github.com/pencil-code/}}
which is a high-order finite-difference code for solving ordinary and
partial differential equations \citep{2021JOSS....6.2807P}.

\subsection{Diagnostics quantities}

The hydrostatic non-convecting solution of \Equs{equ:conti}{equ:entro}
has only a very thin convectively unstable layer near the surface
\citep{Br16} due to the strong temperature and density dependence of
the heat conductivity from the Kramers' law. Thus the Rayleigh number
is quoted from the statistically steady states of the simulations and
defined as
\begin{eqnarray}
\Ra_{\rm t} = \frac{g_{\rm m} R^4}{\nu \chiSGSo} \left(-\frac{1}{\cP} \frac{{\rm d} \brac{s}_{\theta\phi t}}{{\rm d} r}\right)_{\rm m},
\end{eqnarray}
where the subscript m and $\brac{.}_{\theta\phi t}$ refer to
evaluating quantities at $r=0.5R$ and to horizontal and temporal
averaging, respectively. As shown in \cite{2019A&A...631A.122K}, the
SGS entropy diffusion $\chiSGSo$ is the dominant contributor to the
convective stability measured by the Rayleigh number.

The rotational influence on the flow is measured by the Coriolis
number
\begin{eqnarray}
  \Co = \frac{2\Omega_0}{\urms k_R},
\label{equ:Co}
\end{eqnarray}
where $\Omega_0$ is the rotation rate of the star, $\urms$ is the
volume averaged rms velocity within a spherical radius $r<R$, and
where $k_R=2\pi/R$ corresponds to the scale of the largest convective
eddies. To facilitate comparisons to other studies
\citep[e.g.][]{2008ApJ...676.1262B,2020ApJ...902L...3B} an alternative
vorticity-based formulation
\begin{eqnarray}
\Co^{(\omega)} = \frac{2\Omega_0}{\orms}
\end{eqnarray}
is also used, where $\orms$ is the volume averaged rms vorticity
$\bm\omega=\bm\nabla\times\UUU$ within $r < R$. Rossby numbers
corresponding to the Coriolis numbers are given by $\Ro=\Co^{-1}$.

The fluid and magnetic Reynolds numbers are defined as
\begin{eqnarray}
  \Rey = \frac{\urms}{\nu k_R},\ \ \ \ReM = \frac{\urms}{\eta k_R},\label{equ:Rey}
\end{eqnarray}
and the SGS P\'eclet number is given by
\begin{eqnarray}
  \Pe = \frac{\urms}{\chiSGSo k_R}.
\end{eqnarray}

Mean quantities are considered to be azimuthal averages (in
cylindrical coordinates):
\begin{eqnarray}
  \mean{f}(\varpi,z,t) = \frac{1}{2\pi} \int f (\varpi,\phi,z) d\phi,
\end{eqnarray}
where $\varpi=r\sin\theta$ is the cylindrical radius. Temporal
averages over the statistically stationary state are often
additionally applied. Most of the averaged results are presented in
cylindrical coordinates $(\varpi,z)$ but in some cases it is more
relevant to discuss vector quantities transformed to spherical polar
coordinates $(r,\theta,\phi)$. Such quantities are marked with
additional superscript sph, for example, $\UUU^{\rm sph}$.

\section{Results}
\label{sect:results}

Three sets of simulations were made. In the first set (HD) a single
purely hydrodynamic simulation HD1 without rotation was made. In the
second set (RHD), rotation is included with all other parameters
unchanged using thermodynamically equilibrated snapshot from run
HD1. Five rotation rates were explored in the models RHD[1-5]. The
final set of simulations (MHD[1-5]) were branched off from the
corresponding RHD runs with additional weak random seed magnetic
fields. A subset of the MHD runs (MHD[1,3,5]) were remeshed to a
higher resolution (runs MHD[1,3,5]h) and run with lower diffusivities
to test the robustness of the results. This set of runs is referred to
as MHDh. The runs and a number of diagnostics are listed in
\Table{tab:runs1}.

\begin{figure*}[t!]
  \includegraphics[width=\textwidth]{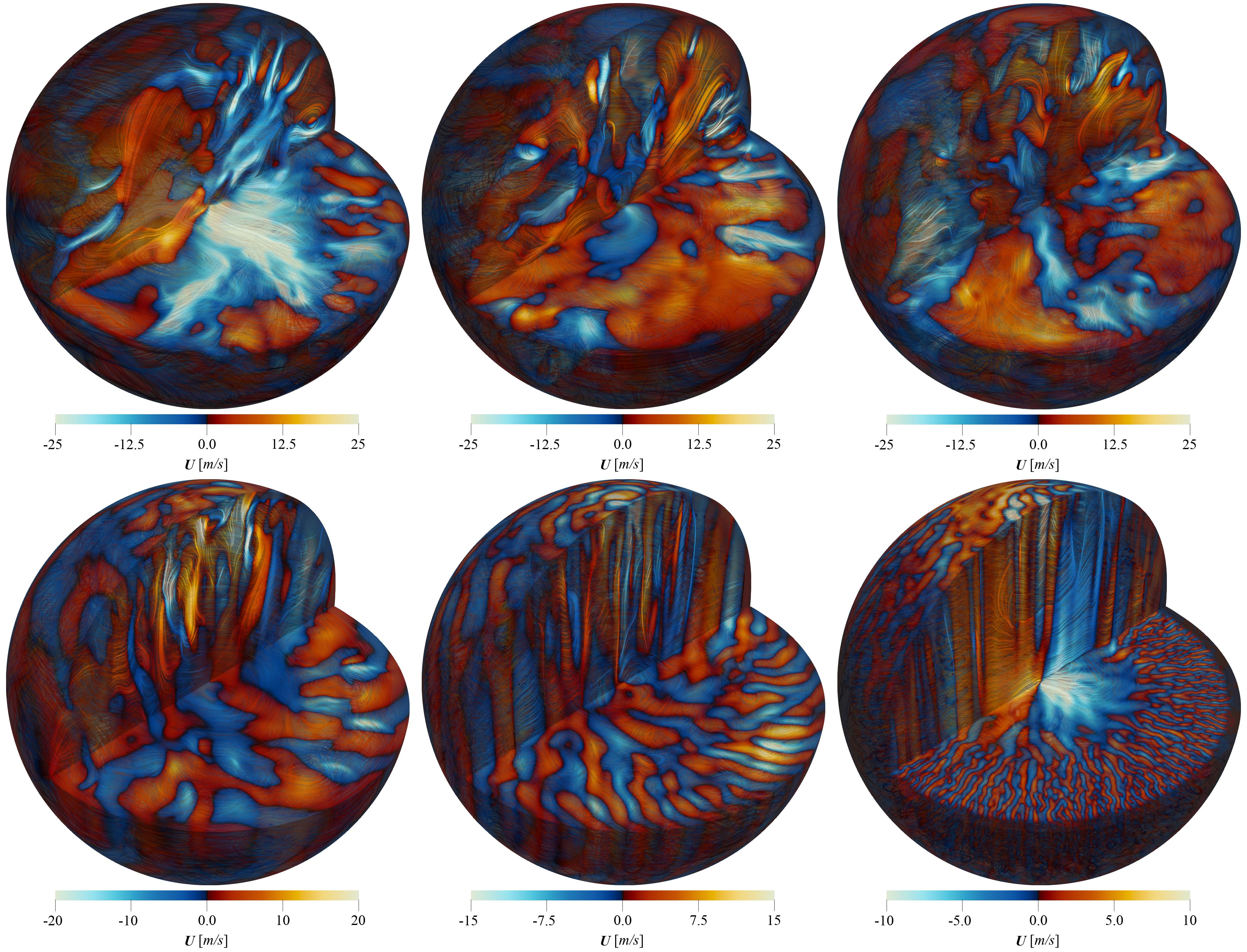}
  \caption{Velocity fields $\UUU^{\rm sph}$ in the Runs~HD1, RHD1,
    RHD2 (top row from left to right), and RHD3, RHD4, and RHD5
    (bottom row). The colour contours indicate radial velocity
    $U_r^{\rm sph}$ and the streamlines are colour-coded according to
    the local value of $U_r^{\rm sph}$.}
\label{fig:uu_sph_288a7}
\end{figure*}

\begin{figure}[t!]
\centering
  \includegraphics[width=.9\columnwidth]{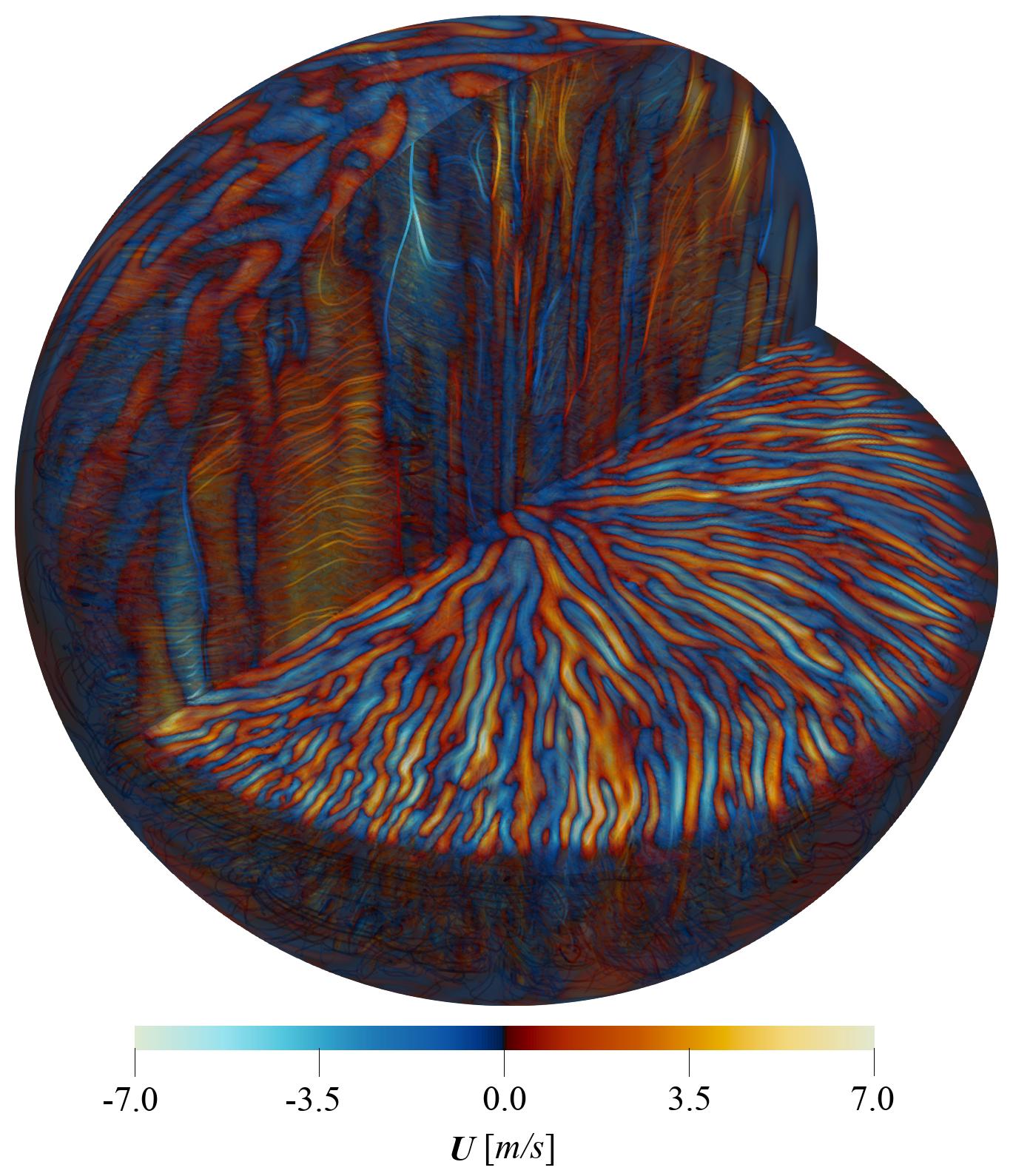}
  \caption{Same as \Figa{fig:uu_sph_288a7} but for run MHD5.}
\label{fig:uu_cart_mhd288a7_Om10b_c1}
\end{figure}

\subsection{Description of convective states}

Representative flows and convection patterns from runs HD1 and
RHD[1-5] are shown in \Figa{fig:uu_sph_288a7}. The non-rotating run
HD1 shows a large-scale dipolar convection cell which fills
essentially the whole star. This is similar to patterns expected for
convection in a sphere near the onset from linear stability analysis
\citep{Ch61}, and to those found in simulations of non- or slowly
rotating partially and fully convective stars in various contexts
\citep[e.g.][]{BP09,2016AdSpR..58.1475O,2020arXiv200108452M}. Rotation
destroys the dipolar convection cell but the flows continue to be
predominantly large scale for runs RHD1 and RHD2 with Coriolis numbers
0.6 and 1.9, respectively. For more rapid rotation the convection
cells begin to show clearer rotational alignment and the horizontal
size of the flow structures decreases, see the lower row of
\Figa{fig:uu_sph_288a7}. This is particularly clear in the two most
rapidly rotating cases RHD4 and RHD5, where the flows are dominated by
helical convection columns that pierce the whole star. In the
hydrodynamic run RHD5, a strong axial jet develops. Similar flow
structures have been recently reported from simulations of Boussinesq
convection in rotating full spheres
\citep[][]{2020arXiv201212061L}. This feature is not present in the
corresponding magnetohydrodynamic run MHD5, see
\Figa{fig:uu_cart_mhd288a7_Om10b_c1}. In the other cases the
differences between hydrodynamic and magnetohydrodynamic runs are less
drastic such the flow patterns are qualitatively unchanged. The
velocities are, however, in general lower in the MHD cases, see the
4th column of \Table{tab:runs1}. Small scale convection near the
surface of the star \citep[e.g.][]{HRY15a} is not captured in the
current simulations because of the modest density stratification in
comparison to real stars. The results of the higher resolution MHDh
runs are very similar to those of the MHD runs which is manifested by
the small differences in $\urms$ and $\brms$ between corresponding
runs, see \Table{tab:runs1}.

\begin{figure}
  \includegraphics[width=\columnwidth]{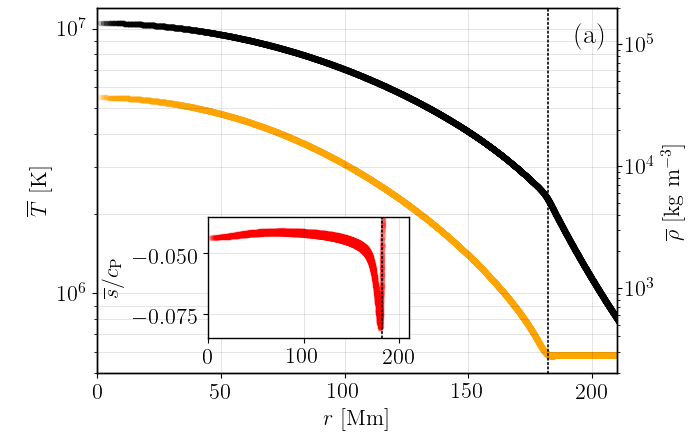}
  \includegraphics[width=\columnwidth]{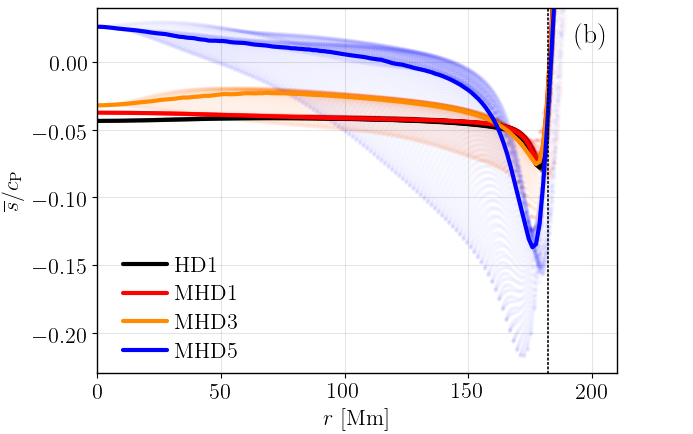}
  \caption{Top panel: Time-averaged radial profiles of azimuthally
    averaged temperature $\mean{T}$ (orange, left axis) and density
    $\mean{\rho}$ (black, right axis) as functions of radius from the
    non-rotating run HD1. The inset shows the radial profile of the
    mean specific entropy $\mean{s}/\cP$. Bottom: mean specific
    entropy $\mean{s}/\cP(r)$ from Runs~HD1, MHD1, MHD3, and MHD5 as
    indicated by the legend. The dots show a scatter plot of
    $\mean{s}(\varpi,\theta)$ as a function of spherical radius
    $r$. The dotted vertical line indicates the upper boundary of the
    convective layer in the non-rotating run HD1.}
\label{fig:pthermo_r_hd288a7}
\end{figure}

The statistically stationary horizontally averaged thermodynamic state
from the non-rotating run HD1 is shown in
\Figa{fig:pthermo_r_hd288a7}(a). The temperature varies by an order of
magnitude within the star whereas the density varies by a factor of
roughly 20. This corresponds to roughly five pressure or three density
scale heights, respectively. In the exterior of the star the
temperature is nearly constant and the density decreases there
exponentially as a function of radius. In comparison to other similar
studies, the density contrast in the current simulations is thus
somewhat greater than in \cite{DSB06}, comparable to that in
\cite{2020ApJ...902L...3B}, and smaller by a factor of a few in
comparison to \cite{2008ApJ...676.1262B} and
\cite{YCMGRPW15,2016ApJ...833L..28Y}.

Panel (b) of \Figa{fig:pthermo_r_hd288a7} shows a scatter plot of the
temporally and azimuthally averaged specific entropy
$\mean{s}(\varpi,z)/\cP$. Remarkably, the mean entropy decreases
toward the centre of the star below $r \lesssim 70$~Mm. Thus roughly
30 per cent of the core of the star is stably stratified according to
the Schwarzschild criterion although the whole star is continuously
mixed by vigorous convection. Similar configurations have recently
been discussed in other contexts where the star has a clearly defined
radiative core
\citep[e.g.][]{2015ApJ...799..142T,2017ApJ...845L..23K,2017ApJ...851...74B,2019GApFD.113..149K}. Such
formally stable, yet convecting, regions are thought to arise due to
cool plumes originating near the surface penetrating deep into the
stellar interior and regions that would otherwise be convectively
stable. This process is sometimes referred to as cool entropy rain
\citep[e.g.][]{Br16}. The idea of such surface-driven stellar
convection goes back to the studies of \cite{SN89} and
\cite{Sp97}. Theoretically, the energy transport in such stably
stratified but still convective layers is explained by a non-local
non-gradient contribution to the heat flux
\citep{1961JAtS...18..540D,De66}. These regions are referred to as
Deardorff zones \citep[see, e.g.][]{Br16,2017ApJ...845L..23K}. The
significance of Deardorff zones in fully convective stars is that they
can have an impact for the large-scale dynamos in that the magnetic
fields in the possibly weakly stably stratified core are less
susceptible to be buoyant.

\Figu{fig:pthermo_r_hd288a7}(b) shows further that in run MHD1 the
convective layer is unstably stratified throughout whereas in run MHD3
the Deardorff layer reappears. The behavior of run MHD1 is puzzling
and possibly related to the strong anti-solar differential rotation
developing in that run. In the most rapidly rotating case, MHD5, the
stratification is again unstable throughout. This can be understood in
terms of an increasing critical Rayleigh number as the rotation
increases such that a steeper temperature gradient is needed to drive
convection \citep[e.g.][]{Ch61}. The current simulations were made
with rather modest Rayleigh numbers (see the 8th column in
\Table{tab:runs1}) and the effects of rotation are likely to be much
weaker in real stars where the Rayleigh numbers are more than ten
orders of magnitude greater. \Figu{fig:pthermo_r_hd288a7}(b) also
shows that the latitudinal variation of mean specific entropy
increases as a function rotation indicated by the larger scatter
around the horizontally averaged means.

\begin{figure}
  \includegraphics[width=\columnwidth]{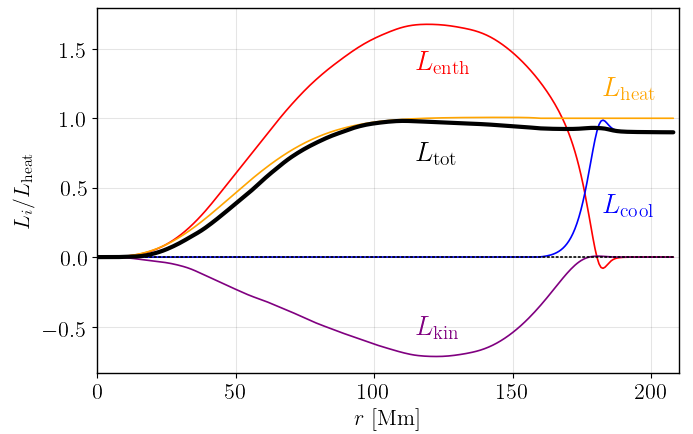}
  \caption{Contributions to the luminosity from enthalpy (red),
    kinetic energy (purple), cooling (blue), and heating (orange)
    fluxes in Run~HD1.}
\label{fig:plot_fluxes_hd288a7_ad}
\end{figure}

The luminosities related to the radiative, enthalpy, kinetic energy,
cooling and heating fluxes are given by
\begin{eqnarray}
  \mLrad  &=& - 4 \pi r^2 \brac{K}_{\theta\phi t} \frac{\pd \brac{T}_{\theta\phi t}}{\pd r},\label{equ:Lrad}\\
  \mLenth &=&  \ \ \ 4 \pi r^2 \cP \brac{(\rho U_r^{\rm sph})' T'}_{\theta\phi t},\\
  \mLkin  &=&  \ \ \ 2 \pi r^2 \brac{\rho \UUU^2 U_r^{\rm sph}}_{\theta\phi t}, \\
  \mLvisc &=& - 8 \pi r^2 \nu \brac{\rho U_i^{\rm sph} \mathsf{S}_{ir}^{\rm sph}}_{\theta\phi t}, \\
  \mLcool &=& -\int_0^{r} 4 \pi r^2 \brac{\mathcal{C}}_{\theta\phi t}\ dr, \\
  \mLheat &=& \ \ \ \int_0^{r} 4 \pi r^2 \brac{\mathcal{H}}_{\theta\phi t}\ dr,\label{equ:Lheat}
\end{eqnarray}
where the primes indicate fluctuations from an azimuthally averaged
mean. Furthermore, the total convected luminosity is given by
\citep{CBTMH91}
\begin{eqnarray}
\mLconv = \mLenth + \mLkin.
\end{eqnarray}
The dominant contributions to the total luminosity from
\Eqs{equ:Lrad}{equ:Lheat} from the non-rotating hydrodynamic run HD1
are shown in \Figa{fig:plot_fluxes_hd288a7_ad}. The luminosities due
to viscous and radiative fluxes are less than one per cent of the
total everywhere and therefore they are not shown. The radiative flux
is likely underestimated in the current simulations due to a value of
$K_0$ that is smaller than that corresponding to an M5 dwarf with the
used $\Lratio$. The effect of this is that the convective velocities
in the simulations are somewhat higher than in the target
star. Assessing the effect of this for the resulting flows and dynamos
is postponed to future studies. The dominant components are due to the
outward enthalpy flux ($\mLenth$) and the inward kinetic energy flux
($\mLkin$). The maximum value of the latter is close to the stellar
luminosity which is compensated by a correspondingly higher flux of
enthalpy such that the total convected luminosity $\mLconv$ matches
the luminosity of the star everywhere except near the surface where
the cooling is effective. In the rotating cases the enthalpy and
kinetic energy luminosity contributions decrease while they still
transport the majority of the flux, similarly as in the fully
convective models of \cite{2020ApJ...902L...3B}.

\begin{figure*}
\centering
  \includegraphics[width=.33\textwidth]{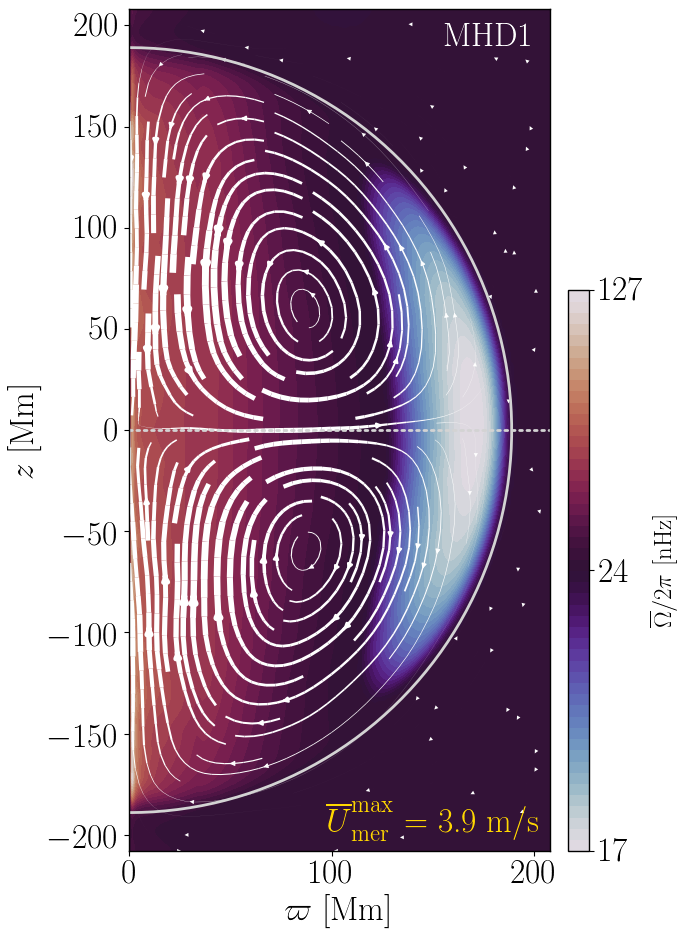}\includegraphics[width=.33\textwidth]{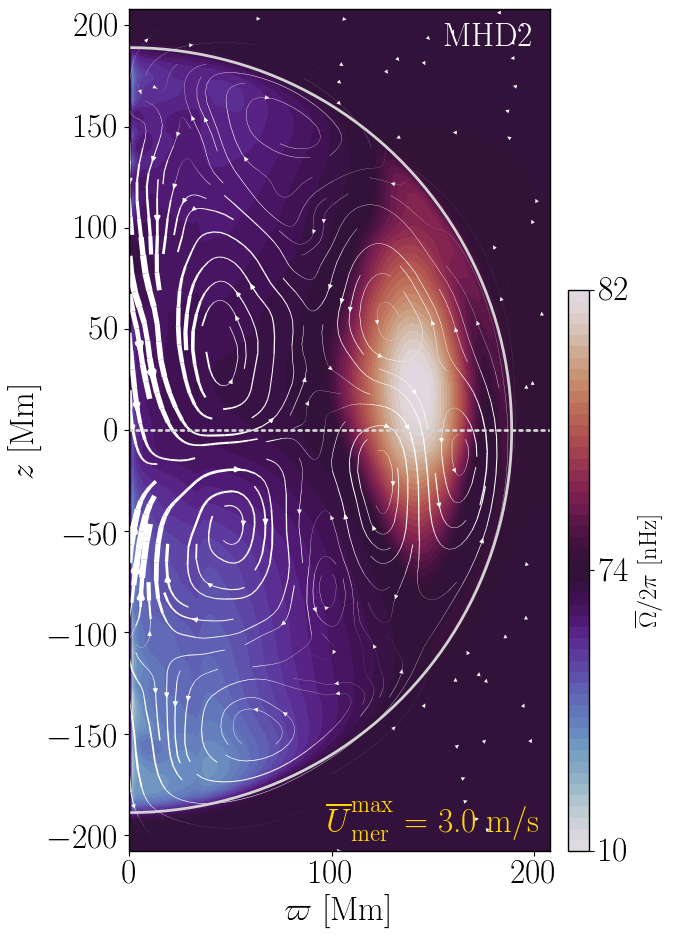}\includegraphics[width=.33\textwidth]{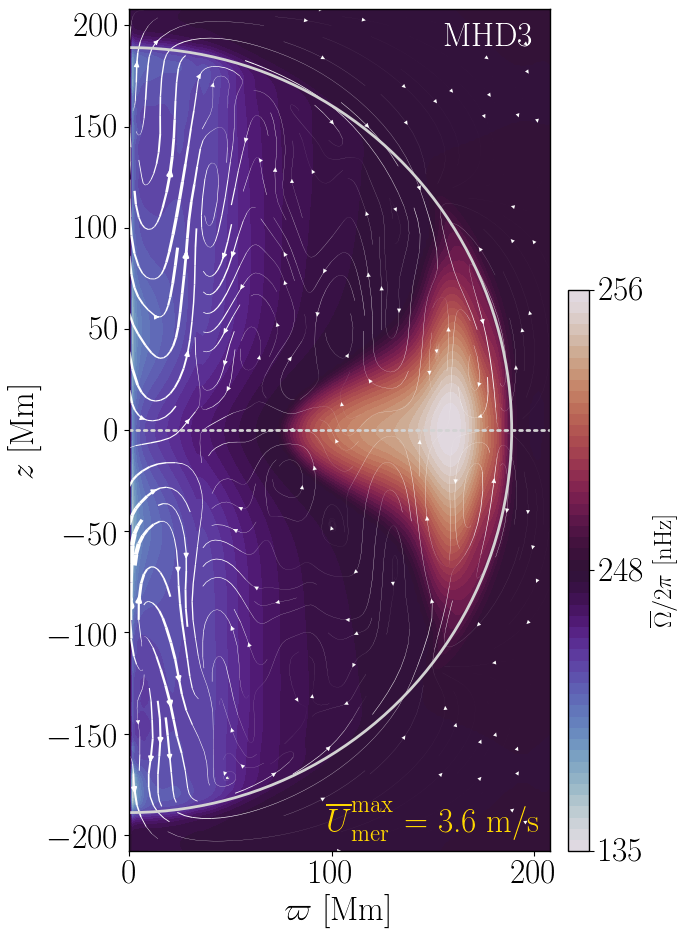}
  \includegraphics[width=.33\textwidth]{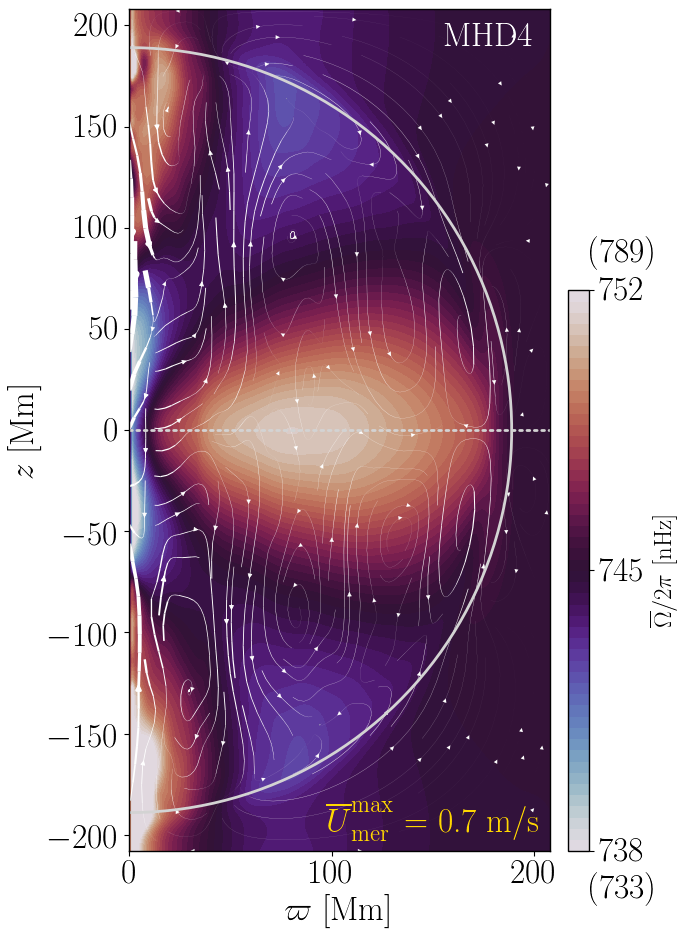}\includegraphics[width=.33\textwidth]{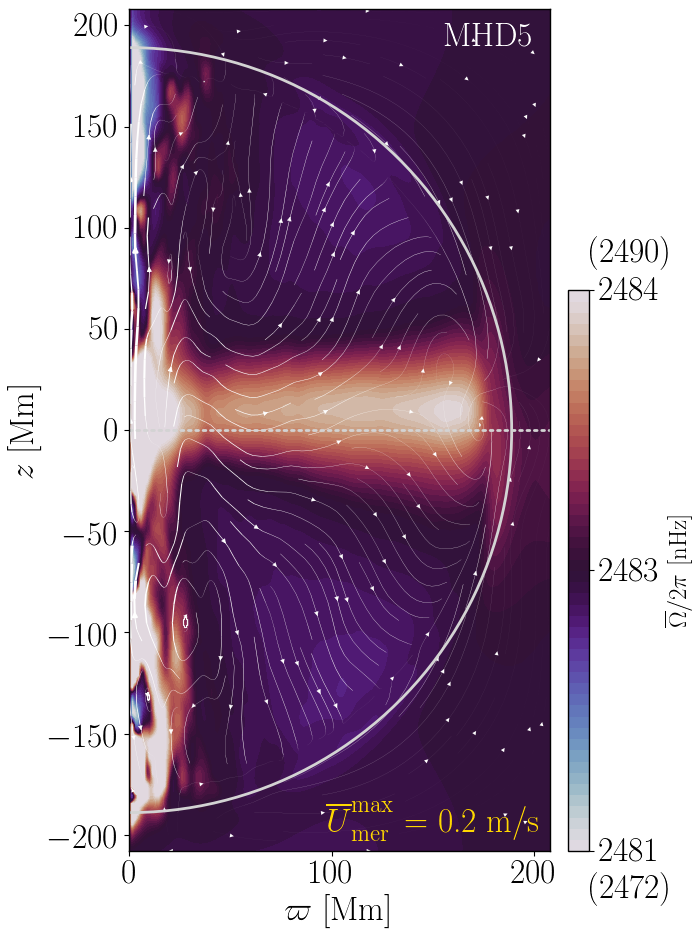}
  \caption{Temporally and azimuthally averaged rotation profiles
    $\mean{\Omega}(\varpi,z)$ for the simulations in set MHD. The
    values for runs MHD4 and MHD5 are clipped to highlight features
    near the equator. The full ranges are indicated in the parenthesis
    above and below the colourbar. The arrows indicated the mass flux
    due to meridional circulation $\mean{\rho}\mUUU_{\rm mer}$. The
    arrow widths are proportional to $|\mean{\rho}\mUUU_{\rm
      mer}|$. The maximum meridional flow speeds are indicated on the
    lower right corners of each panel.}
\label{fig:pOm_mhd288a7}
\end{figure*}

\begin{table}[t]
\centering
\caption[]{Differential rotation parameters.}
  \label{tab:DR}
       \vspace{-0.5cm}
      $$
          \begin{array}{p{0.12\linewidth}ccc}
          \hline
          \hline
          \noalign{\smallskip}
          Run & \Delta_\Omega^{(r)} & \Delta_\Omega^{(\mean{\theta})}(60\degr) & \Delta_\Omega^{(\mean{\theta})}(75\degr) \\
          \hline
          \hline
          RHD1  & -6.56  & -1.18  & -2.75  \\
          RHD2  &  0.15  &  0.24  &  0.45  \\
          RHD3  &  0.30  &  0.15  &  0.30  \\
          RHD4  &  0.087 &  0.018 &  0.050 \\
          RHD5  & -0.012 &  2.6 \cdot 10^{-3}  & -5.8 \cdot 10^{-3} \\
          \hline
          MHD1  & -2.96  & -0.89   & -1.99 \\
          MHD1h & -2.41  & -0.87   & -1.88 \\
          MHD2  &  0.13  &  0.19   &  0.37 \\
          MHD3  &  0.15  &  0.064  &  0.16 \\
          MHD3h &  0.11  &  0.053  &  0.13 \\
          MHD4  & \ \ \ 1.1 \cdot 10^{-3} & 7.5 \cdot 10^{-3}  & 4.2 \cdot 10^{-4}\\
          MHD5  & -8.3 \cdot10^{-5} & 3.0 \cdot 10^{-4}  & 1.7 \cdot 10^{-4}\\
          MHD5h & -2.8 \cdot10^{-4} & 2.7 \cdot 10^{-4}  & 2.2 \cdot 10^{-4}\\
          \hline
          \end{array}
          $$
          \tablefoot{The numbers are computed from temporally and
            azimuthally averaged angular velocity $\mOm(r,\theta)$
            with \Eqs{equ:DRr}{equ:DRt} for each run.}
\end{table}

\subsection{Large-scale flows}

The density-stratified and rotating convective envelopes of fully
convective stars have all the ingredients for generating large-scale
mean flows via the $\Lambda$ effect \citep[e.g.][]{R80,R89,Kap19}.
The ensuing differential rotation and meridional circulation are of
prime importance for large-scale dynamos. Simulations in spherical
shells indicate that two qualitatively different flow regimes exist
depending on the Coriolis number. For slow rotation the differential
rotation is anti-solar, such that the equatorial rotation rate is
slower than the polar one, whereas for rapid rotation a solar-like
rotation profile with equatorial acceleration is obtained
\citep[e.g.][]{Gi77,2007Icar..190..110A,KMB11,GSKM13,GYMRW14}. The
watershed value of the Coriolis number in such simulations is close to
unity \citep[e.g.][]{GYMRW14}.

The averaged rotation rate is given by
\begin{eqnarray}
\mean{\Omega}(\varpi,z) = \Omega_0 + \mean{U}_\phi(\varpi,z) / \varpi.
\end{eqnarray}
Similarly, the averaged meridional flow is given by
\begin{eqnarray}
\mUUU_{\rm mer}(\varpi,z) = (\mean{U}_\varpi, 0, \mean{U}_z),
\end{eqnarray}
where the velocities correspond to those in cylindrical coordinates.
Both $\mOm$ and $\mUUU_{\rm mer}$ are averaged in time over the
statistically steady part of the simulations. The averaging period are
listed in the 12th column of \Table{tab:runs1}. The amplitude of the
differential rotation is measured by the differential rotation
parameters \citep[e.g.][]{KMCWB13}:
\begin{eqnarray}
\Delta_\Omega^{(r)} & = & \frac{\mOm(r_{\rm top},\theta_{\rm eq})-\mOm(r_{\rm
    bot},\theta_{\rm eq})}{\mOm(r_{\rm top},\theta_{\rm eq})},\label{equ:DRr}\\
\Delta_\Omega^{(\mean{\theta})}(\theta) & = &
\frac{\mOm(r_{\rm top},\theta_{\rm eq})-\mOm(r_{\rm top},\mean{\theta})}{\mOm(r_{\rm top},\theta_{\rm eq})},
\label{equ:DRt}
\end{eqnarray}
where $r_{\rm top}=0.9R$, $r_{\rm bot} = 0.1R$, $\theta_{\rm eq} =
0\degr$ corresponds to the equator, and $\mean{\theta}$ refers to an
average of $\mOm$ from latitudes $+\theta$ and
$-\theta$. \Table{tab:DR} summarizes the results for
$\Delta_\Omega^{(r)}$ and $\Delta_\Omega^{(\mean{\theta})}$ with
$\theta=60\degr$ and $75\degr$. The reason not to use the surface
velocities for measuring the differential rotation is that the values
at $r=R$ are likely to be affected by the damping of velocities
applied in the exterior. Similarly, the values near the axis are the
most uncertain and thus off-axis and off-pole values for
$\Delta_\Omega^{(r)}$ and $\Delta_\Omega^{(\mean{\theta})}$ are used.

\begin{table*}[t]
\centering
\caption[]{Volume and time-averaged kinetic and magnetic energy densities.}
  \label{tab:energies}
       \vspace{-0.5cm}
      $$
          \begin{array}{p{0.05\linewidth}ccccccc}
          \hline
          \hline
          \noalign{\smallskip}
          Run & \Ekin [10^5{\rm J}/{\rm m}^3] & \Ekin^{\rm DR}/\Ekin & \Ekin^{\rm MC}/\Ekin & \Emag [10^5{\rm J}/{\rm m}^3] & \Emag/\Ekin & \Emag^{\rm tor}/\Emag & \Emag^{\rm pol}/\Emag \\
          \hline
          HD1   & 71.4  & 0.047  & 0.200  &    -  &    -    &   -    &   -   \\
          \hline
          RHD1  & 68.3  & 0.519  & 0.040  &    -  &    -    &   -    &   -   \\
          RHD2  & 21.5  & 0.179  & 0.045  &    -  &    -    &   -    &   -   \\
          RHD3  & 35.4  & 0.640  & 0.019  &    -  &    -    &   -    &   -   \\
          RHD4  & 15.5  & 0.579  & 0.024  &    -  &    -    &   -    &   -   \\
          RHD5  & 11.4  & 0.127  & 0.018  &    -  &    -    &   -    &   -   \\
          \hline
          MHD1  & 29.6  & 0.306  & 0.072  &  3.58  &  0.121  & 0.095  & 0.080 \\
          MHD1h & 29.6  & 0.267  & 0.077  &  4.06  &  0.137  & 0.072  & 0.074 \\
          MHD2  & 18.9  & 0.140  & 0.045  &  0.51  &  0.027  & 0.074  & 0.068 \\
          MHD3  & 13.8  & 0.395  & 0.025  &  2.37  &  0.172  & 0.256  & 0.063 \\
          MHD3h & 12.0  & 0.297  & 0.025  &  3.25  &  0.271  & 0.169  & 0.053 \\
          MHD4  & 3.01  & 0.117  & 0.014  &  4.83  &  1.604  & 0.044  & 0.053 \\
          MHD5  & 0.63  & 0.016  & 0.014  &  11.1  & 17.528  & 0.017  & 0.039 \\
          MHD5h & 0.63  & 0.014  & 0.015  &  15.5  & 24.730  & 0.022  & 0.057 \\
          \hline
          \end{array}
          $$ \tablefoot{The total kinetic energy density is $\Ekin =
            \onehalf \brac{\rho \UUU^2}$ where the brackets denote
            volume averaging within spherical radius $r \leq R$. The
            corresponding energy densities for the differential
            rotation and meridional circulation are given by
            $\Ekin^{\rm DR} = \onehalf \brac{\rho \mUUp^2}$ and
            $\Ekin^{\rm MC} = \onehalf \brac{\rho (\mUUvp^2 +
              \mUUz^2})$, respectively. The energy densities of the
            total, toroidally averaged toroidal and poloidal magnetic
            fields are given by $\Emag = \brac{\BBB^2/2\mu_0}$,
            $\Emag^{\rm tor} = \brac{\mBBp^2/2\mu_0}$, and $\Emag^{\rm
              pol} = \brac{(\mBBvp^2 + \mBBz^2})/2\mu_0$,
            respectively.}
\end{table*}

\Figu{fig:pOm_mhd288a7} shows the rotation profiles of the runs in the
set MHD. The case with the slowest rotation (MHD1), with
$\Prot=433$~days and $\Co=0.7$, shows anti-solar differential
rotation. This is described by negative values of
$\Delta_\Omega^{(r)}$ and $\Delta_\Omega^{(\mean{\theta})}$, see the
6th row of \Table{tab:DR}.  The anti-solar rotation profile is
associated with a single-cell counter-clockwise meridional flow
pattern with outward flow at the equator. This is similar to runs 2b
and 2c of \cite{DSB06}, see their Fig.~11. The amplitude of the
meridional flow is relatively large, about $4$~m/s which corresponds
to 20 per cent of the overall rms velocity. These results coincide
qualitatively with simulations of thinner convective shells in
spherical coordinates \citep[e.g.][]{KKB14}. In run MHD2, with $\Prot
= 144$~days and $\Co = 2.0$, the rotation profile is solar-like with
positive $\Delta_\Omega^{(r)}$ and $\Delta_\Omega^{(\mean{\theta})}$,
see the 7th row of \Table{tab:DR}. Thus the transition between
anti-solar and solar-like profiles occurs in the Coriolis number range
$0.7 \ldots 2.0$. This is slightly lower than in spherical shell
convection \citep[e.g.][]{KMB11,GYMRW14,2018A&A...616A.160V} where the
transition occurs typically somewhere between $\Co = 2 \ldots
3$. However, due to the differences in the setups including the
geometry, Prandtl numbers, spatial resolution, and Reynolds and
P\'eclet numbers there is no reason to expect that the transition
would occur precisely at the same $\Co$. Nevertheless, it is safe
to conclude that the anti-solar to solar-like transition occurs at a
similar rotational influence in fully and partially convective models.

The amplitudes of the differential rotation are the largest in the
cases MHD1 and MHD2, see 6th and 7th lines of \Table{tab:DR}. A
comparison of $\Delta_\Omega^{(r)}$ and
$\Delta_\Omega^{(\mean{\theta})}$ with \Figa{fig:pOm_mhd288a7}
indicates that the latter somewhat exaggerates the amplitude of the
differential rotation for these runs because the highest values of
$\mOm$ in both cases occur in small regions near the axis where the
accuracy of the $\phi$-averages is the poorest. Nevertheless, the
amplitude of the differential rotation is substantially larger in
these two runs in comparison to the more rapidly rotating cases,
although the fraction of the kinetic energy in the mean flows in the
latter is relatively low, see the 2nd to 4th columns of
\Table{tab:energies}. The rotation profile of the MHD2 is also
asymmetric with respect to the equator. This is most likely caused by
the asymmetric large-scale magnetic field configuration in that
simulation (see below). Similar asymmetric large-scale flow and
magnetic field configurations were reported in \cite{DSB06}, see their
Fig.~11. Furthermore, asymmetry is absent in the corresponding
hydrodynamic run RHD2 (not shown). The meridional flow pattern in run
MHD2 consists of four (two) cells near to (away from) the rotation
axis. The meridional flow is also asymmetric near the equator with
maximum flow speeds of the order of $3$~m/s. Comparison between runs
MHD[1-2] with corresponding hydrodynamic runs RHD[1-2] shows a
reduction of differential rotation of roughly 20 per cent, except for
$\Delta_\Omega^{(r)}$ in run MHD1 which is reduced by more than 50 per
cent.

In run MHD3 the rotation profile is similar to that in MHD2 but no
equatorial asymmetry is present, see \Figa{fig:pOm_mhd288a7}. In MHD3
and the more rapidly rotating cases the maximum meridional circulation
amplitudes occur outside the star and the flows within the star are
significantly weaker. However, as shown in \Figa{fig:pOm_mhd288a7},
the mass flux due to these exterior flows is weak due to the rapidly
decreasing density for $r > R$. The values of $\Delta_\Omega^{(r)}$
and $\Delta_\Omega^{(\mean{\theta})}$ in runs MHD[2-3] are similar to
those of the Sun. However, magnetic quenching of differential rotation
in run MHD3 in comparison to the corresponding hydrodynamic run RHD3
is stronger than for runs MHD2/RHD2. This is likely explained by the
stronger magnetic fields in run MHD3 in comparison to MHD2, see the
5th column of \Table{tab:runs1} and \Table{tab:energies}.

In the two most rapidly rotating runs (MHD4 and MHD5) the amplitude of
the differential rotation decreases further such that in the latter
run the deviations from solid body rotation are very small, see the
lower row of \Figa{fig:pOm_mhd288a7} and rows 11 to 13 in
\Table{tab:DR}. The meridional flow also decreases and the largest
values are always found outside of the star in both cases. The
rotation profile shows axially aligned structures consistent with the
Taylor--Proudman balance in run MHD3 but a similar dominance is not as
clear from the mean flows in cases MHD4 and MHD5. This is possibly
because the mean flows are weak in both cases such that these models
are nearly in solid body rotation. This is expected because for rapid
enough rotation turbulent angular momentum transport, and therefore
the mean flows, are quenched such that the system approaches solid
body rotation for $\Omega_0 \rightarrow \infty$
\citep[e.g.][]{KR93,GYMRW14}. However, inspection of instantaneous
velocity fields reveals flow structures that are very much in line
with Taylor-Proudman constraint, see also
\Figas{fig:uu_sph_288a7}{fig:uu_cart_mhd288a7_Om10b_c1}. The higher
resolution MHDh simulations do not qualitatively differ from their
lower-resolution counterparts. The overall kinetic energy in the MHDh
runs is slightly lower than in the corresponding MHD runs. A similar
trend is seen for $\Delta_\Omega^{(r)}$ and
$\Delta_\Omega^{(\mean{\theta})}(\theta)$ for runs MHD1h and MHD3h.
This is likely due to the stronger magnetic fields in the MHDh runs,
see Tables~\ref{tab:runs1} and \ref{tab:energies}.

\begin{figure*}[t!]
  \includegraphics[width=\textwidth]{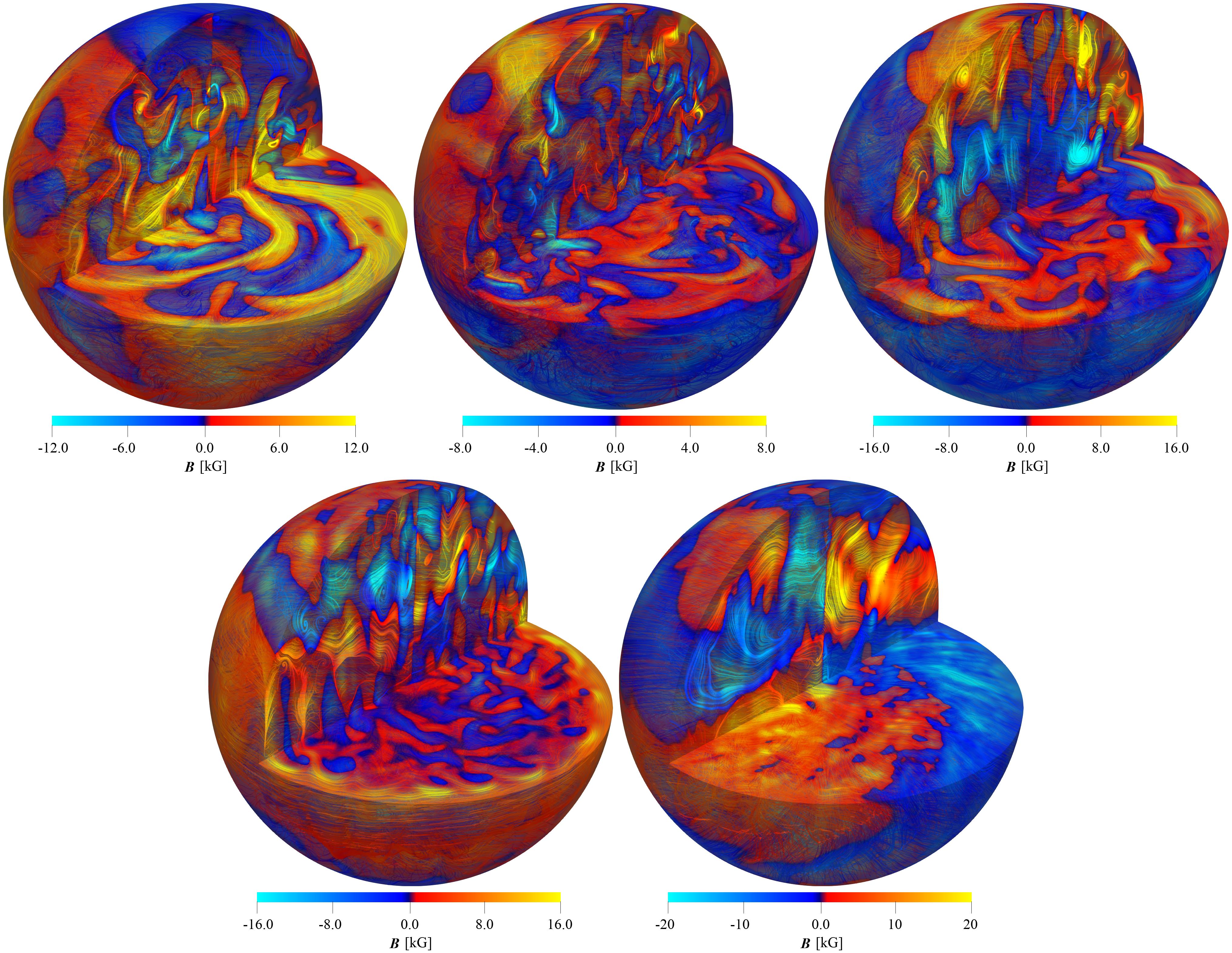}
  \centering
  \caption{Instantaneous magnetic fields $\BBB^{\rm sph}$ in the
    Runs~MHD1, MHD2, MHD3 (top row from left to right), and MHD4, and
    MHD5 (bottom row). The colour contours indicate radial magnetic
    field $B_\phi^{\rm sph}$ and the streamlines are colour-coded
    according to the local value of $B_\phi^{\rm sph}$.}
\label{fig:bb_sph_288a7}
\end{figure*}

The simulations reported in \cite{DSB06} all have anti-solar
differential rotation. The Coriolis numbers were not given in that
study but it is possible to calculate $\Co$ according to \Equ{equ:Co}
with the data presented in the paper.  See \Appendix{sec:comptoDobler}
for more details regarding quantitative comparisons to the study of
\cite{DSB06} and how to transform quantities between code and physical
units.. The Coriolis numbers for their runs 2[b-e] are 0.9, 4.9, 18,
and 47; see \Table{tab:Dobler_runs}. Based on these values it is
somewhat surprising that the differential rotation is anti-solar in
their most rapidly rotating cases. However, the rotation profiles in
\cite{DSB06} are shown from the saturated dynamo regime where the
magnetic field has a considerable back-reaction to the
flow. Furthermore, in the most rapidly rotating cases the
supercriticality of convection is likely to be even weaker than in the
present simulations. Thus it is plausible that the two most rapidly
rapidly rotating cases (2d and 2e) in \cite{DSB06} are comparable to
the current runs MHD4 and MHD5 with weak differential rotation in
general.

The rotation profiles of all runs with solar-like rotation profiles
(MHD[2-5]) all have multiple meridional circulation cells. This is
independent of magnetic fields because the corresponding hydrodynamic
runs have qualitatively similar meridional flow structure. On the
other hand, mean-field models of fully convective stars, relying on
parameterization of turbulence in terms of turbulent viscosity and
$\Lambda$ effect \citep[][]{2017MNRAS.466.3007P,2018ApJ...859...18P}
almost invariably produce a single-cell meridional circulation
pattern. On the other hand, the amplitude of the differential rotation
in the $\Prot=10$~d model of \cite{2017MNRAS.466.3007P} is very close
to that in Run~MHD4.

\begin{figure*}
\centering
  \includegraphics[width=0.5\textwidth]{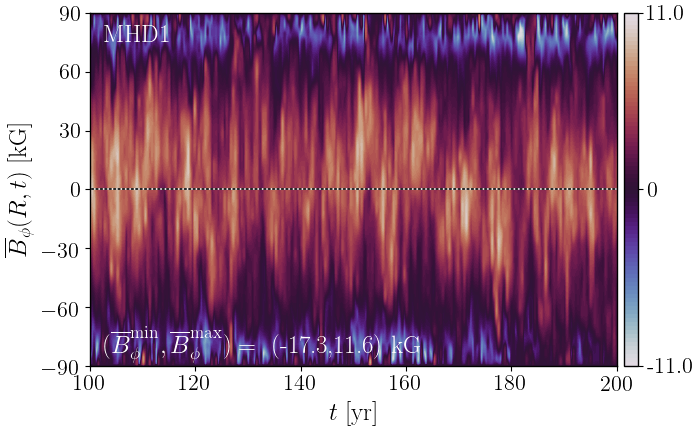}\includegraphics[width=0.5\textwidth]{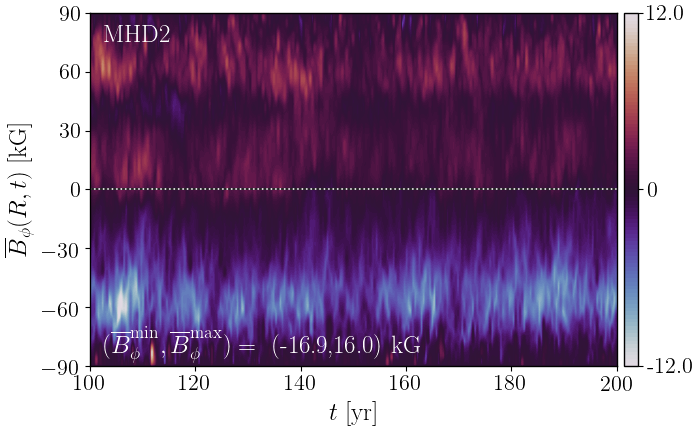}
  \includegraphics[width=0.5\textwidth]{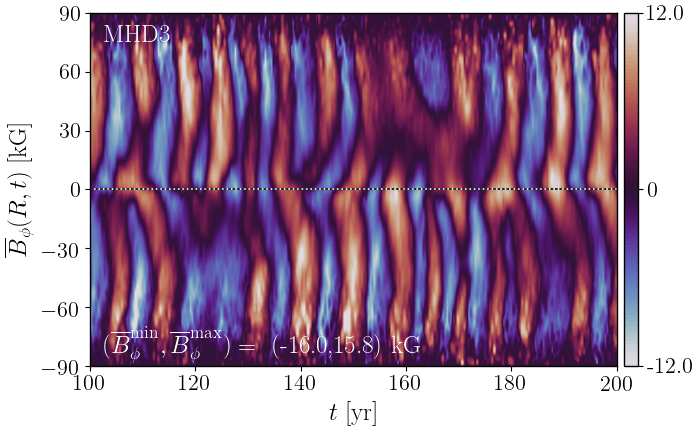}\includegraphics[width=0.5\textwidth]{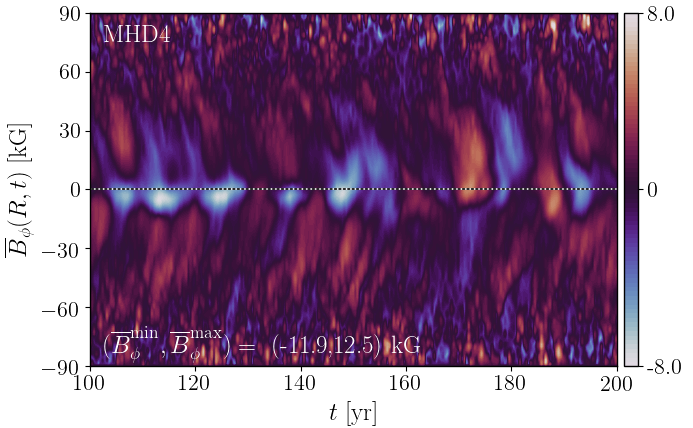}
  \includegraphics[width=0.5\textwidth]{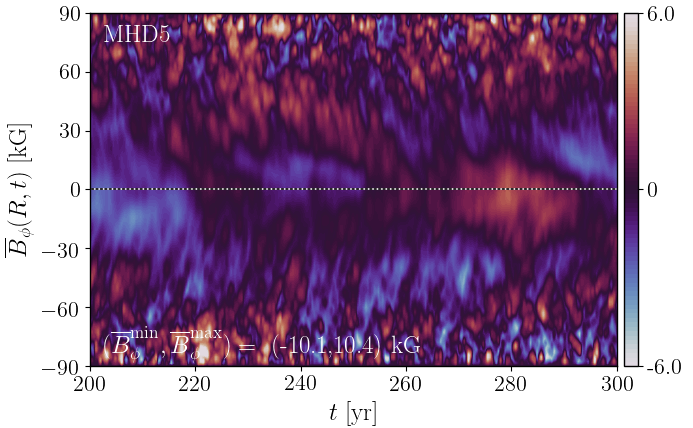}
  \caption{Time-latitude diagrams of the mean toroidal magnetic field
    $\mBBp(R,t)$ for a period of 100~yr for the MHD runs as indicated
    by the legends. The dotted horizontal line indicates the equator.}
\label{fig:pbfly_Bphi_mhd288a7}
\end{figure*}

\subsection{Dynamo solutions}

All of the runs in sets MHD and MHDh produce large-scale dynamos
starting from a low amplitude ($\sim 1$~G) random small-scale seed
field. Representative instantaneous magnetic fields from runs MHD[1-5]
are shown in \Figa{fig:bb_sph_288a7}. While the slowly rotating cases
MHD[1-2] do not appear to show large-scale order at the first glance,
the more ordered large-scale structure of the magnetic field in the
intermediate and rapidly rotating cases MHD[3-5] is immediately
clear. Using the same seed field in the non-rotating run HD1 leads to
a decaying solution such that there is no small-scale dynamo in that
case (run MHD0).

\begin{figure*}
\centering
  \includegraphics[width=0.3\textwidth]{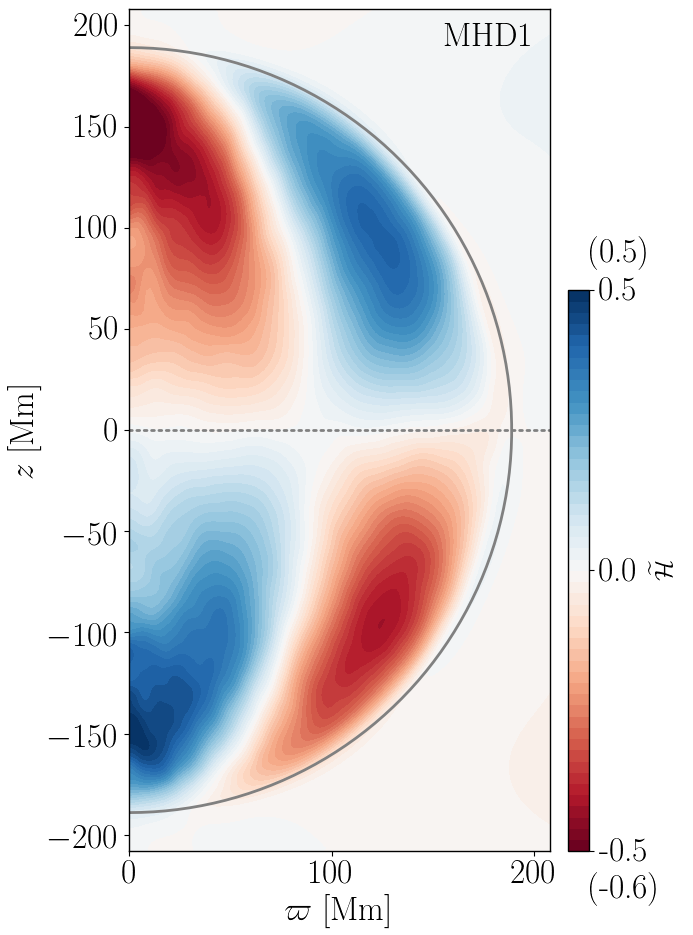}\includegraphics[width=0.3\textwidth]{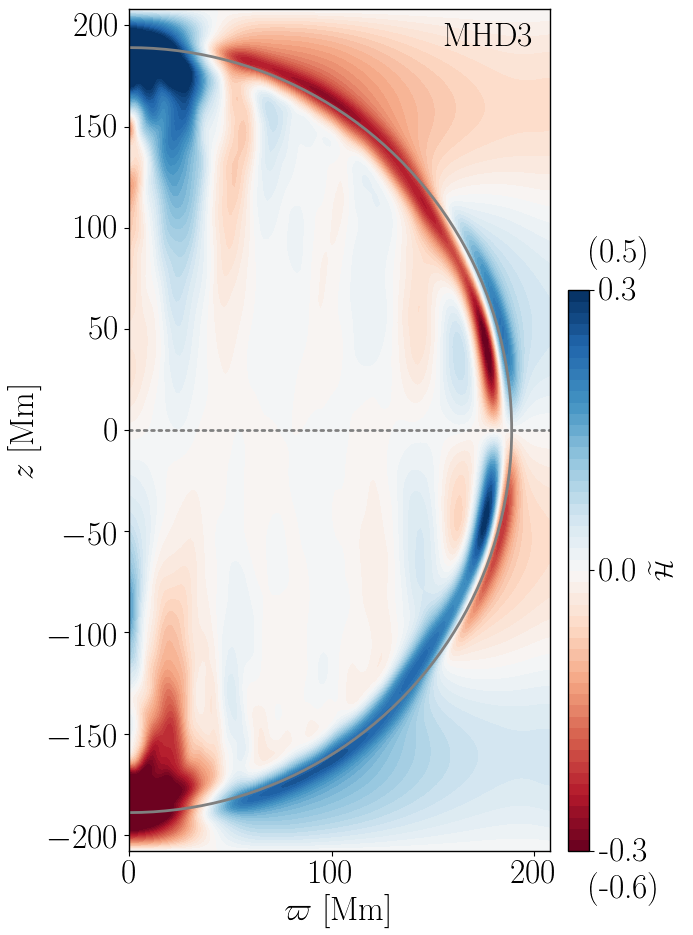}\includegraphics[width=0.3\textwidth]{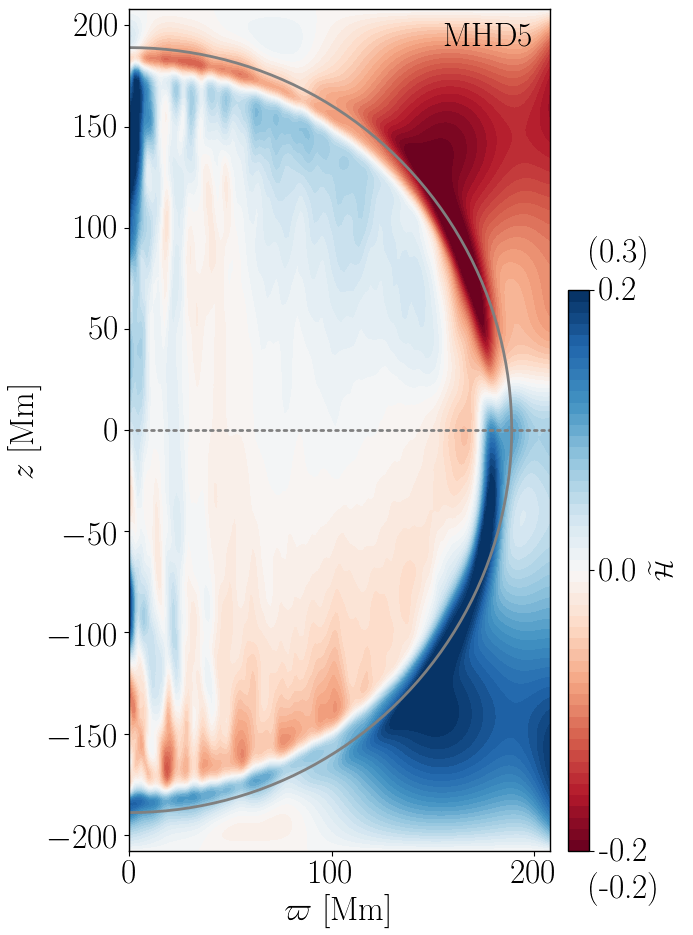}
  \caption{Azimuthally averaged normalized kinetic helicity
    $\tilde{\cal H}(\varpi,z)={\cal H}(\varpi,z)/{\urms \orms}$ for
    runs MHD1 (left), MHD3 (middle), and MHD5 (right). Data ranges are
    clipped for better legibility. The full ranges are indicated
    above and below the colourbars.}
\label{fig:pkinheli}
\end{figure*}

The mean field is considered to be the azimuthally averaged field
$\mBBB(r,\theta,t)$ or $\mBBB(\varpi,z,t)$. Space-time diagrams of
$\mBBp(R,\theta,t)$ for a period of 100 years for each of the MHD runs
are shown in \Figa{fig:pbfly_Bphi_mhd288a7}. The model with the
slowest rotation (MHD1, $\Prot=430$~days) produces a quasi-stationary
large-scale field without sign reversals during the simulated
period. This is similar to results obtained from spherical wedges and
fully spherical shells in the slowly rotating regime where
differential rotation is anti-solar
\citep{2017A&A...599A...4K,2018A&A...616A..72W,2018ApJ...863...35S}. A
similar quasi-stationary solution is also obtained in case MHD2 that
has solar-like differential rotation. Similar quasi-stationary
configurations have been reported also from spherical shells with
solar-like differential rotation \citep[e.g.][]{BBBMT10,GSdGDPKM15}
along with solutions with apparently random sign reversals
\citep[e.g.][]{FF14,HRY16}. This type of solutions appears to be the
norm in a parameter regime where the Coriolis number is just high
enough for solar-like differential rotation to appear
\citep[e.g.][]{2018A&A...616A..72W}. Intriguingly, the kinetic
helicity, ${\cal H} = \mean{\bm\omega \bcdot \UUU}$, in these runs
shows a profile with a mostly positive (negative) values in the upper
part of the CZ in the northern (southern) hemisphere, see the left
panel of \Figa{fig:pkinheli}. This is at odds with the usual
argumentation for helical convective eddies in a stratified atmosphere
for which ${\cal H} < 0$ for $\gggg\bm\cdot\mvOm < 0$.  However,
qualitatively similar results have been reported from a slowly
rotating fully convective setup in the context of proto-neutron stars
\citep{2020arXiv200108452M}.

\begin{figure*}
\centering
  \includegraphics[width=0.25\textwidth]{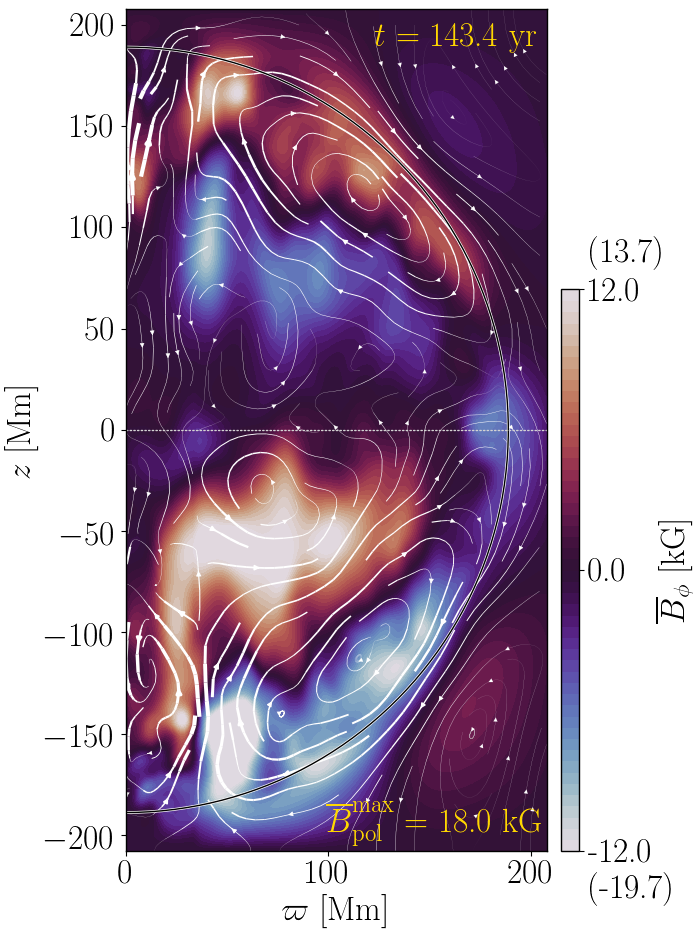}\includegraphics[width=0.25\textwidth]{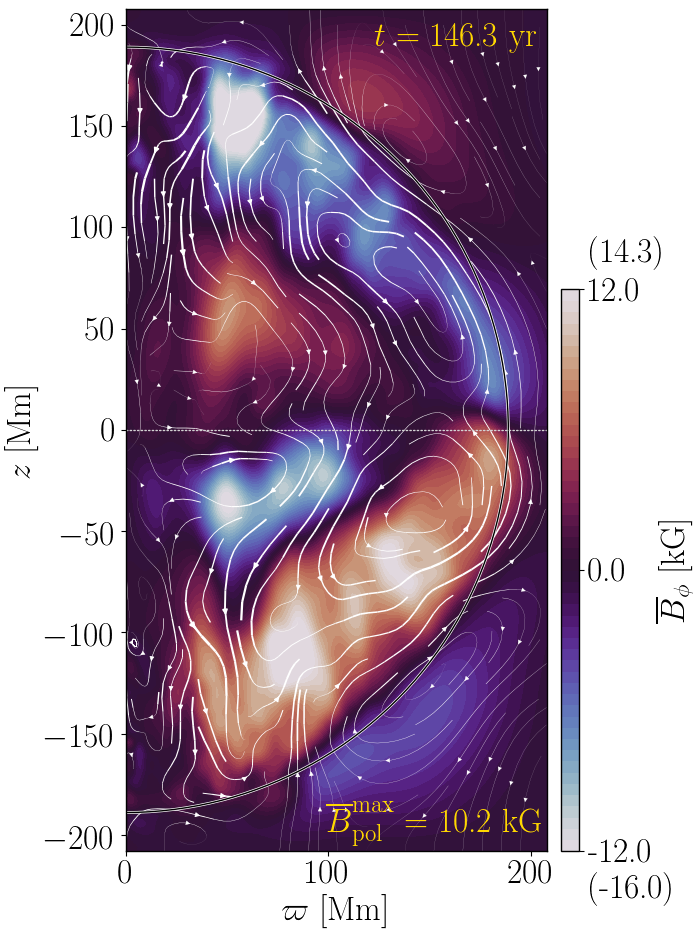}\includegraphics[width=0.25\textwidth]{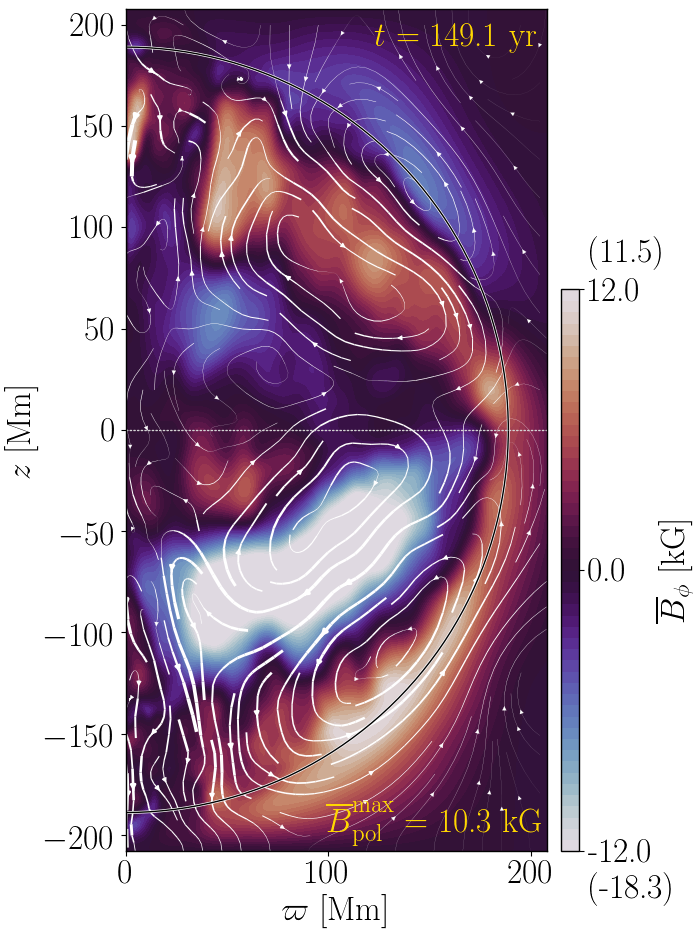}\includegraphics[width=0.25\textwidth]{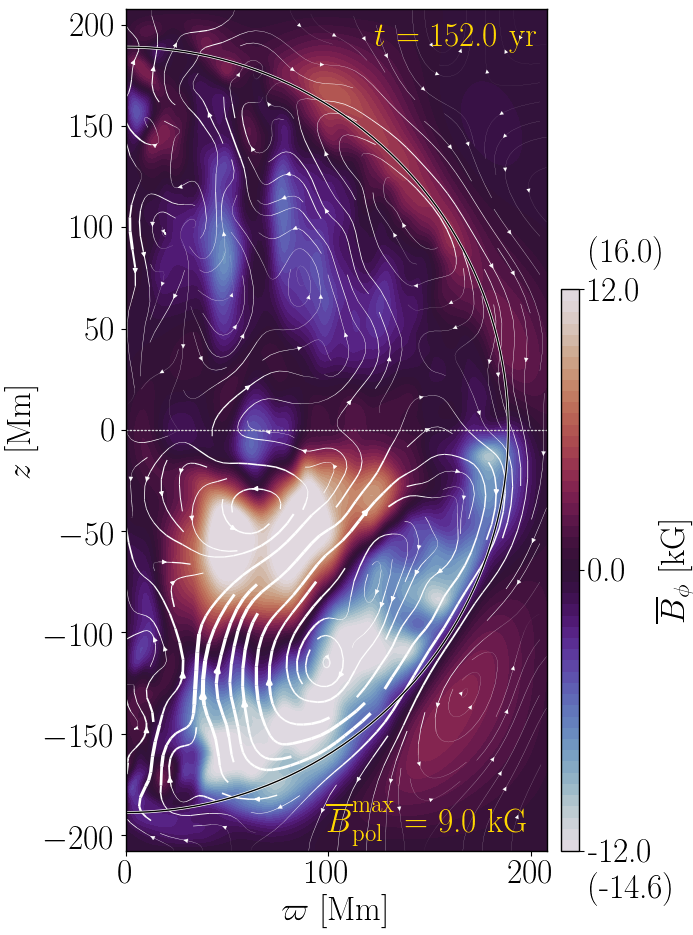}
  \caption{Azimuthally averaged toroidal (colour contours) and
    poloidal (arrows) magnetic fields from four times between
    $t=143.4$ and $152$ years in run MHD3. The width of the arrows is
    proportional to the strength of the poloidal field. The values of
    $\mean{B}_\phi$ are clipped to $\pm12$~kG, and maximum and minimum
    values are quoted in parenthesis above and below the colourbar
    ranges, respectively. The amplitude of the poloidal field is
    indicated on the lower right corner of each panel.}
\label{fig:Bfieldt_MHD3}
\end{figure*}

\begin{figure}
\centering
  \includegraphics[width=0.5\columnwidth]{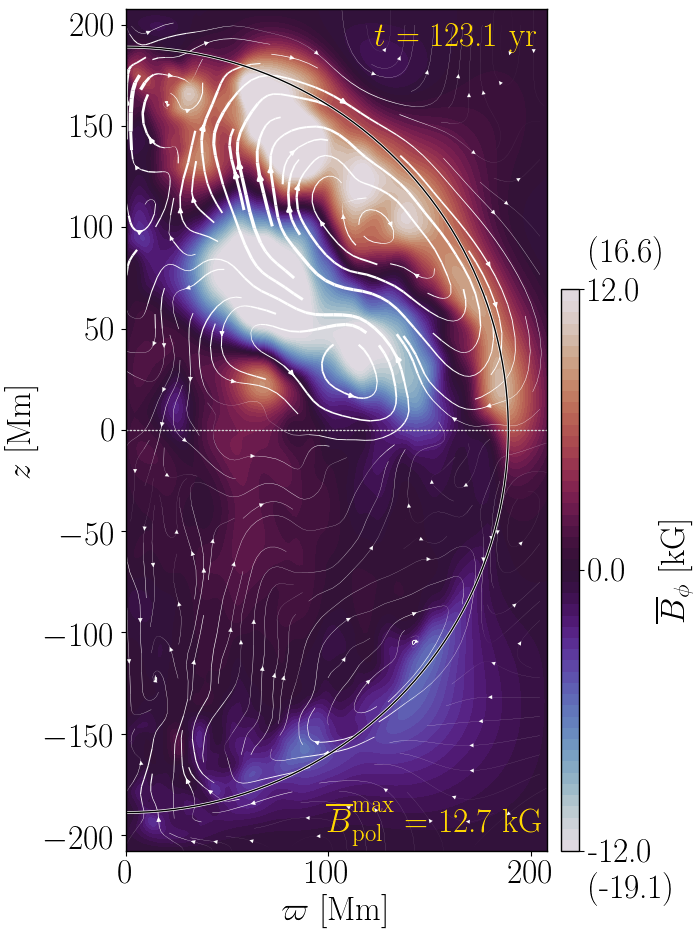}\includegraphics[width=0.5\columnwidth]{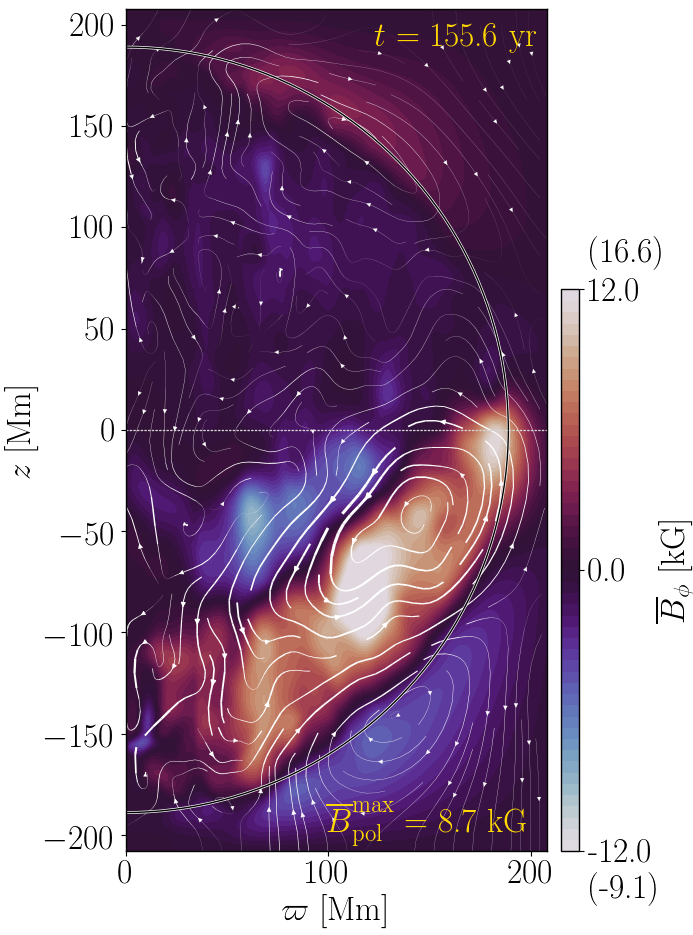}
  \caption{Same as \Figa{fig:Bfieldt_MHD3} but for $t=123.1$ (left) and
    $t=155.6$~years (right).}
\label{fig:Bfieldt_MHD3_304_384}
\end{figure}

\begin{figure*}
  \includegraphics[width=\textwidth]{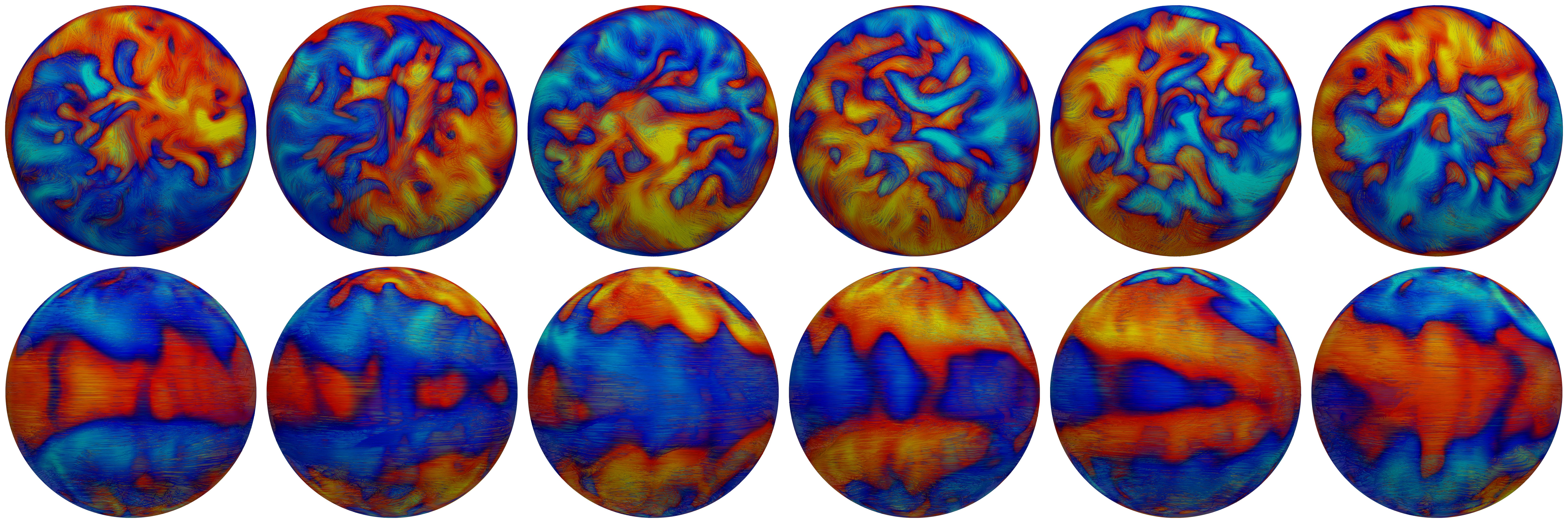}
  \centering
  \includegraphics[width=.4\textwidth]{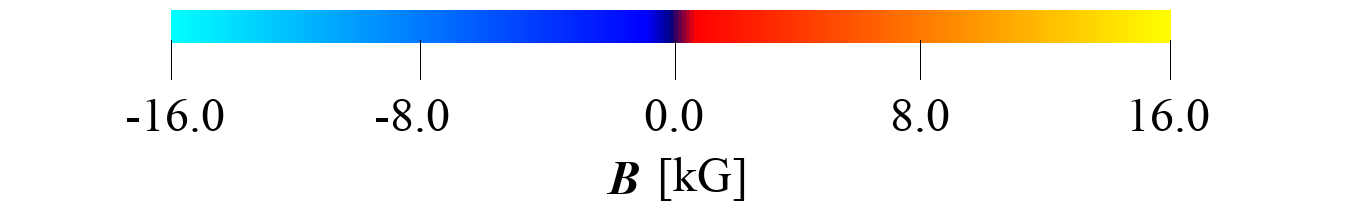}
  \caption{Magnetic field $\BBB^{\rm sph}$ at $r=R$ in run MHD5 from
    six snapshots separated by roughly 6~years each. The top (bottom)
    row shows the magnetic fields pole-on (equator-on). The colour
    contours and the colour-coding of the field lines indicate
    strength of the local radial field. The colour scale is clipped at
    $\pm16$~kG and the extrema of $B_r^{\rm sph}$ are $\pm70$~kG.}
\label{fig:bb_poleequ_mhd288a7_Om10b}
\end{figure*}

In the intermediate rotation regime (run~MHD3), with $\Prot=43$~days
and $\Co = 9.1$, a predominantly axisymmetric and oscillatory
solution is found. In this regime the rotation profile of the star is
solar-like such that the radial gradient of $\mean{\Omega}$ is
positive almost everywhere. In combination with a mostly negative
kinetic helicity on the northern hemisphere the expectation from a
simple $\alpha\Omega$ dynamo is that a poleward-propagating dynamo
wave appears \citep{Pa55b,Yo75}. This is consistent with the outcome
of the simulation, see the left panel in the middle row of
\Figa{fig:pbfly_Bphi_mhd288a7}, although the region where ${\cal H}<0$
is present only near the surface of the star (middle panel of
\Figa{fig:pkinheli}). However, the results for the large-scale
magnetic fields are in agreement with comparable simulations in
spherical shells
\citep[e.g.][]{Gi83,KKBMT10,BMBBT11,2015ApJ...810...80S}. The
evolution of the azimuthally averaged mean magnetic field in the
course of the cycle is shown in \Figa{fig:Bfieldt_MHD3}. The amplitude
of the mean toroidal magnetic field is typically larger than the
corresponding poloidal field but the difference is not particularly
large, and in some instances the maximum poloidal field exceeds the
maximum toroidal field. However, the volume- and time-averaged
toroidal field energy is roughly four times greater than the energy of
the poloidal field, see \Table{tab:energies}. Thus the interpretation
in terms of an $\alpha\Omega$ dynamo remains plausible. However, a
more detailed description, for example, in terms of mean-field
modeling with turbulent transport coefficients derived from the 3D
simulations with the test-field method
\citep[e.g.][]{SRSRC07,2018A&A...609A..51W,2020A&A...642A..66W} is
needed to discern the origin of the large-scale fields. This, however,
is not within the scope of the present study.

In a recent study, \cite{2016ApJ...833L..28Y} presented a spherical
shell simulation extending from $r=0.1R$ to $R$ in a parameter regime
targeting specifically the slowly rotating M dwarf Proxima Centauri
with $\Prot \approx 90$~days. These authors found a similar
poleward-propagating dynamo wave as in the current run MHD3 with a
half-cycle period of roughly 9 years. The run MHD3 with $\Prot =
43$~days has an average half-cycle of around 4-5 years. Although the
parameters of the current simulations and those of
\cite{2016ApJ...833L..28Y} do not fully overlap, the comparison
suggests that neglecting a small core region at the centre of the star
does not play a crucial role for the dynamo solution. Furthermore,
these numerical findings are close to recent observational results
that suggest an activity cycle of about 7 years for Proxima Centauri
\citep{2016A&A...595A..12S,2017MNRAS.464.3281W,2021MNRAS.500.1844K}.

In another recent study, \cite{2020ApJ...902L...3B} reported on a
simulation of a fully convective star in the rapidly rotating
regime. These authors quote a vorticity-based Rossby number $\Ro =
|\bm\nabla\times\uuu|/2\Omega \approx 0.3$ and a velocity-based Rossby
number $\Ro_u = |\uuu|/2\Omega R \approx 0.01$, where
$\uuu=\UUU-\mUUU$ is the fluctuating velocity, for this
simulation. The inverses of the velocity- and vorticity-based Coriolis
numbers of the current study, $1/\Co$ and $1/\Co^{(\omega)}$, are not
exactly the same as the Rossby numbers of \cite{2020ApJ...902L...3B}
because they are based on the total velocity $\UUU$ and thus
overestimate the Rossby number by some factor. Thus the $\Ro$ of the
simulation presented \cite{2020ApJ...902L...3B} is most likely
somewhere between the current runs MHD3 and MHD4 where
$1/\Co^{(\omega)} = 0.27$ and $1/\Co^{(\omega)} = 0.83$.  The ratio
$\Emag/\Ekin$ is also comparable to that quoted by
\cite{2020ApJ...902L...3B} ($0.25$) in runs MHD3 ($0.17$) and MHD3h
($0.27$).

These authors found a dynamo solution which is predominantly
hemispheric. Furthermore, they explain that similar solutions are
found in the parameter regime in the vicinity of the reported
simulation. The current simulations do not indicate a similar behavior
but sweeping conclusions cannot be made based on the very limited
number of simulations done so far. There are some periods in run MHD3
where the magnetic field is predominantly hemispherical (see,
\Figa{fig:Bfieldt_MHD3_304_384}) but such configurations cover only a
small fraction of the total duration of the simulation. Furthermore,
the mean magnetic field in run MHD2 is consistently stronger on the
southern hemisphere (see \Figa{fig:pbfly_Bphi_mhd288a7}). This is
perhaps suggesting that oscillatory hemispheric solutions could also
be realized for parameters not far away from those in the present runs
MHD2 and MHD3. Another example of a possibly hemispheric dynamo was
presented by \cite{2008ApJ...676.1262B} with velocity
fluctuation-based Rossby number of $\approx 10^{-2}$, but there is not
enough information regarding the time evolution of the magnetic fields
to ascertain this definitively (see their Fig.~7 panels b and c).

For the rapid rotators (runs MHD4 and MHD5) the axisymmetric magnetic
fields show no clear cycles while the mean fields appear to migrate
predominantly toward the equator
(\Figa{fig:pbfly_Bphi_mhd288a7}). However, the fraction of the
axisymmetric mean field energy in comparison to the total field energy
is sharply decreasing in runs MHD[4-5] in comparison to intermediate
rotation cases, see the 7th and 8th columns of
\Table{tab:energies}. This is, however, associated with a strong
increase of the total magnetic energy (5ht column of
\Table{tab:energies}), and a closer inspection indicates that the
magnetic fields in theses cases have a strong non-axisymmetric
contribution with a dominant $m=1$ mode, see representative results
from MHD5 in \Figa{fig:bb_poleequ_mhd288a7_Om10b}. This large-scale
non-axisymmetric structure also drifts with respect to the rotating
frame of the star. Such azimuthal dynamo waves were predicted by
mean-field theory \citep[][]{KR80} and have more recently been
reported from simulations of stellar dynamos
\citep[e.g.][]{KMCWB13,CKMB14,YGCR15,2018A&A...616A.160V,2021arXiv210211110N}. In
run MHD4 (MHD5) the $m=1$ magnetic structure drifts in the retrograde
direction with a period of roughly 10 (40) years. These periods are
also similar to those in simulations of rapidly rotating solar-like
stars \citep[e.g.][]{CKMB14,2018A&A...616A.160V}.

The rightmost panel of \Figa{fig:pkinheli} shows that the kinetic
helicity is mostly negative on the northern hemisphere in run MHD5
although the normalized values are smaller than in the less rapidly
rotating cases. Due to the almost totally absent differential
rotation, the most plausible dynamo mechanism in this case is an
$\alpha^2$ dynamo. However, more detailed information of the turbulent
transport coefficients are needed to confirm this.

The higher resolution runs in the MHDh set of simulations do not show
qualitative differences from the corresponding lower resolution
simulations MHD[1,3,5]. The overall magnetic field strength is higher
in the MHDh runs by 15 (MHD1h) to 40 (MHD5h) per cent. Nevertheless,
the spatial structure and temporal evolution of the large-scale
magnetic is similar to those in the lower resolution runs in set
MHD. Therefore the results appear to be robust although higher
resolution studies are still needed to study effects such as efficient
small-scale dynamo action and larger density stratifications.

The rms field strengths in the current runs exceed 10~kG in all cases
except MHD2, see the 5th column of \Table{tab:runs1}. Furthermore, the
order of magnitude of the magnetic energies ($10^5$~J/m$^3$, see the
5th column of \Table{tab:energies}) in the current runs agrees with
\cite{2008ApJ...676.1262B} suggesting similar field strengths and that
the results are robust irrespective of the numerical method. Magnetic
fields exceeding 10~kG are somewhat higher than those observed in M
dwarfs where typical total field strengths are of the order of a few
kilogauss \citep[e.g.][and references
  therein]{2021A&ARv..29....1K}. There are several possible reasons
for the apparent discrepancy, including too low resolution of the
simulations or the observations. A possible way forward is to produce
synthetic observations, such as coarse-grained maps of surface fields
\citep[e.g.][]{YCMGRPW15} or time series corresponding to brightness
variations, from the simulation results and comparing those with
actual observations. This will be addressed in future work.

\section{Conclusions}

An updated version of the star-in-a-box model originally published by
\cite{DSB06} was used to compute hydrodynamic and magnetohydrodynamic
models of a $0.2M_\odot$ fully convective M5 star with various
rotation rates. The current results indicate that the transition from
anti-solar to solar-like differential rotation in fully convective
stars occurs at a comparable rotational regime, as measured by a
Coriolis number, as in simulations of partially convective stars
\citep[e.g.][]{GYMRW14,KKB14}. Furthermore, the dynamo-generated
large-scale magnetic fields are predominantly axisymmetric and
quasi-stationary for slow rotation and axisymmetric and cyclic for
intermediate rotation. For rapid rotation the large-scale fields
become increasingly non-axisymmetric with a dominating low-order
($m=1$) mode. These large-scale non-axisymmetric fields also exhibit
azimuthal dynamo waves where the large-scale magnetic structure drifts
in longitude with little regard of the underlying flows
\citep[e.g.][]{KR80,CKMB14}. Remarkably, very similar transitions at
comparable Coriolis numbers, have been reported from spherical shell
simulations \citep[e.g.][]{KMCWB13,2018A&A...616A.160V}. This suggests
that the dynamos in fully convective stars are fundamentally the same
as in partially convective stars.

The current exploratory set of simulations presented in the current
study is, however, by no means exhaustive. This is illustrated by the
lack of bistable dipolar and multipolar solutions
\citep[e.g.][]{2013A&A...549L...5G,YCMGRPW15} and hemispheric dynamos
\citep{2020ApJ...902L...3B} in the current set of models. A
particularly interesting parameter regime is that of intermediate
rotation, in the vicinity of the current run MHD3 with
$\Prot=43$~days, where cyclic dynamo solutions appear. Similar results
have been reported from anelastic simulations in spherical shells
\citep{2016ApJ...833L..28Y} and in full spheres
\citep{2020ApJ...902L...3B}. This similarity of solutions between very
different numerical approaches is encouraging. Furthermore, this is an
interesting regime also observationally in view of the recent results
on the possible activity cycles of Proxima Centauri and Ross 128 with
rotation periods of the order of 100 days. Further simulations and
comparisons to observations are clearly called for in this range of
parameters.

Another aspect that makes fully convective stars interesting is that
they are often found in post-common envelope binaries (PCEBs)
\citep[e.g.][]{2013MNRAS.429..256P}. The periods of many of these
binaries show residual signals that indicate orbital evolution of the
system. Such signals can be caused by planets although explaining
their existence in such evolved systems is a problem in
itself. Another explanation is that the secondary star hosts a cyclic
dynamo that affects the gravitational quadrupole moment of the star
which is reflected in orbit variations. This mechanism was first
introduced by \cite{1992ApJ...385..621A} and more recent ideas along
similar lines have decreased for example energetic requirements for
orbit variations due to non-axisymmetric large-scale fields
\citep{2020MNRAS.491.1820L}. Recent numerical simulations of
\cite{2020MNRAS.491.1043N,2021arXiv210211110N} demonstrated the effect
of dynamo-generated magnetism on the gravitational quadrupole moment
for partially convective rapidly rotating solar-like stars. Extending
the study of the stellar quadrupole moment to fully convective stars
is a logical step to follow in future publications.

\begin{acknowledgements}
  I thank the anonymous referee, Axel Brandenburg, Fabio del Sordo,
  Oleg Kochukhov, Felipe Navarrete, Carolina Ortiz, G\"unther
  R\"udiger, Dominik Schleicher, and J\"orn Warnecke for fruitful
  discussions and useful comments on the manuscript. I acknowledge the
  hospitality of Nordita during my visit in February-March 2020. The
  simulations were made using the HLRN-IV supercomputers Emmy and Lise
  hosted by the North German Supercomputing Alliance (HLRN) in
  G\"ottingen and Berlin, Germany. This work was supported by the
  Deutsche Forschungsgemeinschaft Heisenberg programme (grant No.\ KA
  4825/2-1).
\end{acknowledgements}

\bibliographystyle{aa}
\bibliography{paper}

\appendix

\section{Comparison to \cite{DSB06}}
\label{sec:comptoDobler}

Since the simulations of \cite{DSB06} were made with the same model as
in the current study, it is possible to make direct quantitative
comparisons of the parameters and results. Using the values of ${\cal
  L}$ and other parameters quoted in Table~1 of \cite{DSB06} it is
possible to compute the outputs of the model in physical units. A
number of such outputs, along with a few dimensionless numbers
transformed to the same definitions as in the present study are listed
in \Table{tab:Dobler_runs}.

The conversion factor between the rotation rates in the
  simulation and the target star was given in \Equ{equ:Omegaconv}. For
  all of the runs of \cite{DSB06} except 1c this yields
\begin{eqnarray}
\Omssi = 12.5 \left(\frac{\gsi}{\Rsi} \right)^{1/2}, 
\end{eqnarray}
for the solar rotation rate $\Omega_\odot = 2.7\cdot
10^{-6}$~s$^{-1}$. For their run 1c the prefactor is 10.1 instead of
12.5. In the most rapidly rotating run of \cite{DSB06} the rotation
rate is $\Omega_0 = 10 (\gsi/\Rsi)^{1/2}$, where $\gsi=\Rsi=1$ in code
units. This means that the rotation period in that run is still more
than 30 days, see the 4th and 5th columns of
\Table{tab:Dobler_runs}. Nevertheless, this run is in the rapidly
rotating regime because of the overall lower convective velocities
than in the present study, compare the 4th column of \Table{tab:runs1}
with the 6th colunm of \Table{tab:Dobler_runs}. This is most clearly
seen in their non-rotating case 2a where $\urms = 9.3~{\rm m/s}$
whereas $\urms =18~{\rm m/s}$ in the current run HD1. Such lower
velocities are likely due to the more efficient radiative energy
transport in their runs. That is, convection transports only a
fraction of the energy instead of practically the whole luminosity as
in the present study.

The conversion of velocities and magnetic fields from simulation to
physical units is done through \citep[cf.\ Appendix~A
  of][]{2020GApFD.114....8K}
\begin{eqnarray}
\UUU^{\rm phys} &=& \left( \frac{\Omega_\odot R}{\Omssi R^{\rm sim}} \right) \UUU^{\rm sim}, \\
\BBB^{\rm phys} &=& \left[ \frac{\mu_0 \rho_0 (\Omega_\odot R)^2}{\mu_0^{\rm sim} \rho_0^{\rm sim} (\Omssi R^{\rm sim})^2} \right]^{1/2} \BBB^{\rm sim},
\end{eqnarray}
where $\mu_0^{\rm sim}=\rho_0^{\rm sim}=1$ are used in the code, and
where the solar rotation rate is used as a reference value.

The Reynolds numbers in \cite{DSB06} were computed using the stellar
radius $R$ as the length scale whereas the corresponding wave number
$k_R = R/2\pi$ is used in the definitions in the present study. Thus
the fluid and magnetic Reynolds numbers with the current definitions
in \Equ{equ:Rey} read:
\begin{eqnarray}
\Rey = \frac{\Rey^{\rm DSB}}{2\pi},\ \ \ReM = \frac{\ReM^{\rm DSB}}{2\pi},
\end{eqnarray}
where the superscript DSB refers to the definitions in
\cite{DSB06}. Furthermore, the P\'eclet number can be computed using
the quantities given in Table~1 of \cite{DSB06}. Note that a constant
$\chi$ was used by \cite{DSB06} and no SGS entropy diffusion is
added. Finally, the Coriolis number according to \Equ{equ:Co} is given
by:
\begin{eqnarray}
\Co = \frac{\Omega_0 R^2}{\nu \pi \Rey_{\rm DSB}}.
\end{eqnarray}
Even though the rotation rates of the runs in \cite{DSB06} are lower,
the Coriolis numbers are comparable to those in the present study. The
fluid Reynolds numbers are generally smaller by a factor of roughly
three while the magnetic Reynolds numbers are somewhat higher in
\cite{DSB06}. This explained by the higher magnetic Prandtl number
($\Pm=2$) in comparison to the present study ($\Pm=0.5$). This also
explains why a small-scale dynamo is excited in run 2a of \cite{DSB06}
with $\ReM=93$ while it is absent in the current run MHD0 with an
almost identical magnetic Reynolds number ($\ReM=95$) \citep[see,
  e.g.][]{SICMPY07,2018AN....339..127K}.

\begin{table*}[t]
\centering
\caption[]{Summary of the runs in \cite{DSB06}.}
  \label{tab:Dobler_runs}
       \vspace{-0.5cm}
      $$
          \begin{array}{p{0.05\linewidth}ccccccccccccc}
          \hline
          \hline
          \noalign{\smallskip}
          Run & {\cal L} & \Lratio & \Omega_0 & P_{\rm rot} [\mbox{days}] & \Co & \urms [{\rm m}/{\rm s}] & \brms [\mbox{kG}] & \nu\ [10^6{\rm m}^2~{\rm s}^{-1}] & \chi\ [10^6{\rm m}^2~{\rm s}^{-1}] & \Rey & \ReM  & \Pe \\
          \hline
          1a  &  0.02  & 8.3\cdot10^{11} &  0.2  &  1594  &   0.4  &  6.6  &  3.5  &  4.5  &  3.0  &  43  &   87  & 65 \\
          1b  &  0.02  & 8.3\cdot10^{11} &  0.2  &  1594  &   0.4  &  7.4  &  4.9  &  3.0  &  2.3  &  62  &  123  & 82 \\
          1c  &  0.01  & 4.2\cdot10^{11} &  0.2  &  1265  &   0.5  &  6.6  &  5.0  &  3.8  &  2.9  &  52  &  103  & 69 \\
          \hline
          2a  &  0.02  & 8.3\cdot10^{11} &  0.0  &    -   &    -   &  9.3  &  3.1  &  6.0  &  6.0  &  46  &   93  & 46 \\
          2b &   0.02  & 8.3\cdot10^{11} &  0.5  &   638  &  0.9   &  7.4  &  8.0  &  6.0  &  6.0  &  37  &   74  & 37 \\
          2c  &  0.02  & 8.3\cdot10^{11} &  2.0  &   159  &  4.9   &  5.2  &   12  &  6.0  &  6.0  &  26  &   51  & 26 \\
          2d  &  0.02  & 8.3\cdot10^{11} &  5.0  &    64  &   18   &  3.5 &    17  &  6.0  &  6.0  &  17  &   35  & 17 \\
          2e  &  0.02  & 8.3\cdot10^{11} &  10   &    32  &   47   &  2.7 &    18  &  6.0  &  6.0  &  14  &   27  & 14 \\
          \hline
          \end{array}
          $$ \tablefoot{All runs have $\Pm=2$. The rms velocity refers
            to the saturated value ($\urms^{\rm sat}$) where such
            value is available and to the kinematic one ($\urms^{\rm
              kin}$) otherwise.}
\end{table*}

\end{document}